\documentclass{aa}
\usepackage{graphicx}
\usepackage{amssymb}
\usepackage{natbib}
\bibpunct{(}{)}{;}{a}{}{,}

\begin{document}

\title{On the importance of the few most massive stars: \\
the ionizing cluster of \object{NGC 588}}

\titlerunning{The ionizing cluster of \object{NGC 588}}

\author{L. Jamet\inst{1} \and E. P\'erez\inst{2} \and M. Cervi\~no\inst{2} \and
G. Stasi\'nska\inst{1} \and R.M. Gonz\'alez Delgado\inst{2} \and J.M.
V{\'\i}lchez\inst{2}}

\offprints{L. Jamet, \email{Luc.Jamet@obspm.fr}}

\institute{LUTH, Observatoire de Meudon, 5 place Jules Janssen, 92195 Meudon
Cedex, France \and Instituto de Astrof{\'\i}sica de Andaluc{\'\i}a (CSIC),
Apartado 3004, 18080 Granada, Spain}


\date{Received / Accepted }

\abstract{We present the results of a double analysis of the ionizing cluster
in \object{NGC 588}, a giant H{\sc II} region (GHR) in the outskirts of the
nearby galaxy \object{M33}. For this purpose, we obtained ground based
long-slit spectroscopy and combined it with archival ground based and space
borne imaging and spectroscopy, in the wavelength range 1100--9800 \AA. A first
modeling of the cluster was performed using integrated properties, such as the
spectral energy distribution (SED), broad band colors, nebular emission
H$\beta$ equivalent width, the main ultraviolet resonance lines, and the
presence of Wolf-Rayet star features. By applying standard assumptions about
the initial mass function (IMF), we were unable to fit satisfactorily these
observational data. This contradictory result led us to carry out a second
modeling, based on a resolved photometric analysis of individual stars in
Hubble Space Telescope (HST) images, by means of finding the best fit isochrone
in color-magnitude diagrams (CMD), and assigning a theoretical SED to each
individual star. The overall SED of the cluster, obtained by integrating the
individual stellar SEDs, is found to fit better the observed SED than the best
solution found through the integrated first analysis, but at a significantly
later stage of evolution of the cluster of 4.2 Myr, as obtained from the best
fit to the CMD. A comparative analysis of both methods traces the different
results to the effects of statistical fluctuations in the upper end of the IMF,
which are significant in \object{NGC 588}, with a computed cluster mass of 5600
M$_{\sun}$, as predicted by \cite{cvl02}. We discuss the results in terms of
the strong influence of the few most massive stars, six in the case of
\object{NGC 588}, that dominate the overall SED and, in particular, the
ionizing far ultraviolet range beyond the Lyman limit.

}

\maketitle

\section{Introduction}

Despite their small number in comparison with Sun-like stars, massive stars are
most influential in their host galaxies. They are responsible for the release
in the interstellar medium of most of the mechanical energy, metals such as
oxygen, neon, sulfur, etc. and possibly carbon, and hard radiation. However,
their formation is still poorly understood. It is generally admitted that
massive stars form mainly in instantaneous starbursts, with a power-law initial
mass function (IMF) with universal slope \citep[e.g.,][]{m03}, although
\cite{s98} estimates that this slope may comprise a natural scatter of a few
tenths between different clusters. Most massive stars are observed and studied
in very massive clusters embedded in giant H{\sc II} regions (GHRs), in which
the ionized gas provides information on the intensity of the (unobservable)
stellar Lyman continuum, via the equivalent width of the nebular H$\beta$ line.
Inversely, modeling accurately a GHR from its emission spectrum requires a good
knowledge of the ionizing spectrum of its source, which is one of the elements
that control the temperature and ionization state of the gas. This ionizing
spectrum is unobservable, and can only be derived from a model of the stellar
source.

The stellar content of a GHR can be determined in two ways: either by observing
the individual stars, or making assumptions regarding the distribution of their
initial masses. The first solution can be used to study the cluster itself via
color-magnitude diagrams \citep[CMDs, e.g. ][]{mwp96} or to synthesize an
ionizing spectrum to be input to photoionization models \citep{rpb02}. However,
such a procedure requires quite high resolution and sensitivity for
extragalactic objects, which can in general be reached only with the HST. In the
vast majority of other cases, only integrated properties of the cluster can be
used, and an IMF, generally a continuous power law, must be supposed.
Nevertheless, \cite{cvl02} showed that if the total initial mass of a cluster
is less than $\sim10^4$ M$_{\sun}$, statistical fluctuations around the mean
IMF can strongly affect the main diagnostics of the cluster, like the H$\beta$
equivalent width $W({\rm H}\beta)$, as well as the determined ionizing spectrum
\citep[see also][]{clp03}. The effects of these fluctuations need to be tested
with observational data.

In this article, we show that the cluster embedded in \object{NGC 588} nicely
illustrates the issue of IMF statistical sampling. Section~\ref{obsdat}
describes all the data sets used, spectroscopic and photometric, ground based
and space borne, either obtained by us or retrieved from different archives.
Further data processing is detailed in Section~\ref{datproc}.
Section~\ref{classi} describes a first modeling of the star cluster, using a
``classical'' approach, aimed at fitting its integrated spectral properties
with a model assuming an analytical IMF. In Section~\ref{statmod} we introduce
the importance of modeling the effects of statistical fluctuations. In
Section~\ref{photappr}, we measure the fluxes of the individual stars detected
in HST images, and use the photometric results to analyze the stellar content
of the cluster, by assigning a model SED to each individual star. The results
from the two methods of modeling are significantly discrepant, and in
Section~\ref{discuss} we explore how this discrepancy can be attributed to the
role of the few most massive stars, whose number is heavily affected by
statistical fluctuations in the case of the low mass cluster ionizing
\object{NGC 588}.

\section{Observational data and standard reduction}
\label{obsdat}

We have performed ground based optical spectroscopic observations of
\object{NGC 588} and have retrieved the available archival spectroscopic and
imaging data sets from the Hubble Space Telescope (HST), the International
Ultraviolet Explorer (IUE), and the Isaac Newton Group (ING).
Table~\ref{tab_obs} contains a summary of the different data sets used in this
study.

\begin{table}
\caption{Journal of Observations.}
\label{tab_obs}
\begin{tabular}{ccc}
\multicolumn{3}{l}{\bf Spectroscopy} \\
\hline
\multicolumn{3}{l}{CAHA (PA=121$^{\circ}$, width=1.2$\arcsec$, 1998-08-27/28)}\\
\hline
range: $\lambda$ ({\AA}) & $\Delta\lambda$ ({\AA}/pixel) & exposure (s) \\
B : 3595--5225 & 0.81 & $4\times1800$ \\
R1: 5505--7695 & 1.08 & $2\times1800$ \\
R2: 7605--9805 & 1.08 & $2\times1800$ \\
\hline
\multicolumn{3}{l}{HST/STIS (PA=-154$^{\circ}$, width=0.2$\arcsec$,
2000-07-13)} \\
\hline
range ({\AA}) & exposure (s) & data set \\
1447--3305 & 1440 & O5CO05010 \\
1075--1775 &  480 & O5CO05020 \\
1075--1775 & 2820 & O5CO05030 \\
\hline
\multicolumn{3}{l}{IUE (PA=102$^{\circ}$, large aperture, 1980-10-03)} \\
\hline
range &  exposure (s) & data ID \\
1150--1980 & 18000 & SWP10273 \\
1850--3350 & 18000 & LWR08943 \\
\hline
\hline
\multicolumn{3}{l}{\bf Imaging} \\
\hline
\multicolumn{3}{l}{HST/WFPC2 (Proposal: 5384, 1994-10-29)} \\
\hline
Filter & exposure (s) & data set \\
F170W & $2\times350$ & U2C60701T/702T \\
F336W & $2\times160$ & U2C60703T/704T \\
F439W & $2\times180$ & U2C60801T/802T \\
F547M & $2\times100$ & U2C60803T/804T \\
\hline
\multicolumn{3}{l}{JKT (1990-08-20/21)} \\
\hline
Band      & Filter & exposure (s)      \\
H$\beta$  & 4861/54 & $2\times1800$    \\
H$\alpha$ & 6563/53 & 1800             \\
H$\alpha$ continuum & 6834/51 & 1300   \\
\hline
\end{tabular}
\end{table}

\subsection{Optical spectra}

Optical spectra were obtained in 1998 August 27/28 with the 3.5m telescope at
the Centro Astron\'omico Hispano Alem\'an (CAHA, Calar Alto, Almer\'\i a). The
observations were made with the dual beam TWIN spectrograph, using the blue and
red arms simultaneously, with the beam-splitter at 5500 {\AA}, and SITe-CCDs of
$2000\times800$ 15$\mu$ pixels. Four exposures of 1800 s each were taken with
the blue arm in the range 3595--5225 {\AA}, while two 1800 s exposures in the
range 5505--7695 {\AA} and two 1800 s in the range 7605--9805 {\AA} were taken
in the red arm (hereafter, the B, R1 and R2 ranges, respectively). The gratings
T7 in the blue and T4 in the red give dispersions of 0.81 and 1.08 \AA/pixel in
second and first spectral order, respectively. A slit, oriented at
PA$=121^{\circ}$ (Fig.~\ref{n588_b}), of size
$240\arcsec\times(1.2\pm0.2)\arcsec$ provided a spectral resolution, similar in
both arms, of FWHM$\sim2$ {\AA}, as measured in the background sky emission
lines.

These spectra were reduced, following the standard procedure, by means of the
IRAF\footnote{IRAF is distributed by the National Optical Astronomy
Observatory, which is operated by the Association of Universities for Research
in Astronomy, Inc., under cooperative agreement with the National Science
Foundation.} package \texttt{noao.twodspec.longslit}. Photometric calibration
was achieved by means of three standard stars, \object{BD+28~4211},
\object{G191-B2B} and \object{GD71}, observed during the same night and setup,
but through a wide slit. We estimated the resulting photometric accuracy to be
3\% in color, except for the $\lambda<4000$ {\AA} range (residuals are $\sim
5$\% intrinsic to this range, $\sim 15$\% with respect to the global spectrum),
where the sensitivity curve drops, and for the $\lambda>8000$ {\AA} range
(accuracy $\sim 30$\%), which is strongly affected by telluric absorption
bands. The absolute photometric level was determined with $\sim 10$\% accuracy.
The reduction included the subtraction of the background measured in the slit
aside the nebula, and consequently, the light of the underlying stellar
population of \object{M33} was removed from our spectra.

\subsection{IUE spectra}
We retrieved two spectra from the final archive IUE Newly Extracted Spectra
\citep[INES\footnote{The INES System has been developed by the ESA IUE Project
at VILSPA. Data and access software are distributed and maintained by INTA
through the INES Principal Center at LAEFF. {\tt
http://ines.laeff.esa.es/}.}:][]{rgs99,cag00,gcs00}. The SWP and LWR spectra
cover a total range of 1150--3350 {\AA}, and were obtained in the
low-resolution mode ($\sim$5 {\AA}) through the large aperture
($9.5\arcsec\times 22\arcsec$; cf. Fig.~\ref{n588_b}).

The two IUE spectra were retrieved from the archive in two formats: the
integrated, fully calibrated spectra, and the two-dimensional frames,
linearized in wavelength and position. The slit position and mean angular
resolution were known with an accuracy of a few arcseconds, a scale comparable
to that of the densest spatial structures of the cluster, and we re-determined
them more accurately following the procedure detailed in \S~\ref{spareg}.
Furthermore, due to significant uncertainties in the photometry of the LWR
spectrum, we limited our use of IUE data to the SWP spectrum.

\subsection{Ground-based images}
We downloaded a set of images of \object{NGC 588} from the Isaac Newton Group
of telescopes archive. These images were taken with the 1m Jacobus Kapteyn
Telescope (JKT) at the Observatorio del Roque de los Muchachos (La Palma),
through the filters H$\alpha$+[N{\sc II}], H$\beta$, and a continuum narrow
band filter close to H$\alpha$.

\begin{figure}
\resizebox{\hsize}{!}{\includegraphics{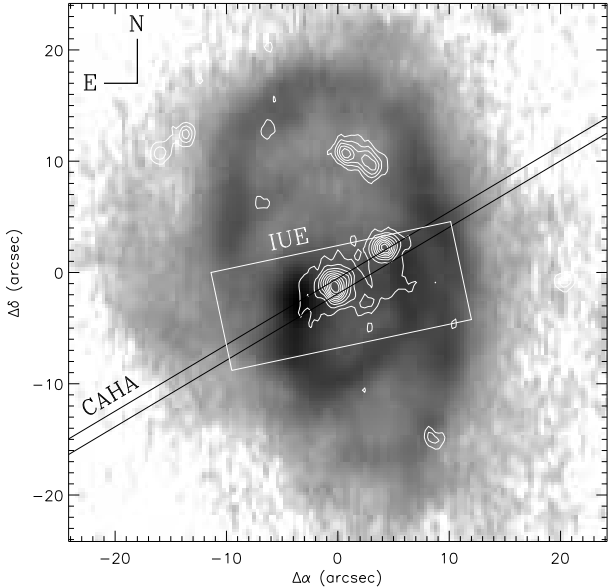}}
\caption{H$\alpha$+[N{\sc II}] image of \object{NGC 588} and CAHA and IUE
slits. The contours represent the H$\alpha$ continuum.}
\label{n588ha}
\end{figure}

Line emission only images were obtained by subtracting a version of the
continuum image scaled with stars common to the three filter images. Then, we
divided the H$\alpha$+[N{\sc II}] image by the H$\beta$ one, thus establishing
an extinction map of the nebula (the intensity of the [N{\sc II}]
$\lambda$6548+6583 doublet being small compared to H$\alpha$, according to our
spectra). We observed no significant gradient of extinction across the object.
We also registered the images with respect to the HST images, in order to
locate the slits of the different spectrographs over the nebula
(Fig.~\ref{n588ha}).

\subsection{HST images}
\label{hstimg}
We retrieved five pairs of WFPC2 images from the HST archive, taken through the
filters F547M, F439W, F336W, F170W and F469N. The images were calibrated with
the on-the-fly calibration system. Each pair of exposures within a filter was
combined with {\tt crrej} in order to perform cosmic ray rejection.

The images were corrected for filter contamination, as recommended by
\cite{mmw02}. These corrections consisted of increasing the measured fluxes by
0.044 magnitude for filter F170W, 0.006, for F336W and 0.002, for F439W. We
also accounted for charge-transfer efficiency loss \citep{whc99}. This was
necessary because of the low background levels in our images ($\sim$0.3 to 0.5
DN). In the case of individual stellar photometry (Section~\ref{stelphot}), we
applied directly the formulae of \cite{whc99} to correct the number of counts
in each star. In the case of integrated fluxes (Section~\ref{classi}), we used
an average correction for each filter based on the rough locus of the cluster
on the images ($X\sim 400$ and $Y\sim 450$) and on the approximate counts found
in the stars that dominate the total cluster flux. These corrections are of
0.037, 0.035, 0.036 and 0.044 magnitude for filters F547M, F439W, F336W and
F170W, respectively. We neglected the effects of geometric distortions, because
the cluster is close to the center of the images, and the photometric effects
of these distortions are small compared to other errors.

We computed the barycentric wavelengths of the filters. For this, we multiplied
the spectral response of each filter by a few SEDs characteristic of young OB
associations, and calculated the corresponding barycentric wavelengths. For
F547M, F439W, F336W, F170W and F469N, we respectively found 5470, 4300, 3330,
1740 and 4690 {\AA}. These wavelengths were then used to compute the values of
the extinction laws to be applied.

\subsection{STIS spectra}
The HST archive contains three STIS spectra through \object{NGC 588}, covering
the two wavelength ranges 1447--3305 {\AA} (with the grating G230L) and
1075-1775 {\AA} (with the grating G140L), the two spectra covering the latter
range having a resolution of $\sim$1.5 {\AA}. We retrieved these spectra with
on-the-fly calibration. The spectra pass only through the brightest star of
\object{NGC 588} (Fig.~\ref{n588_b}), and we used them to identify its spectral
class (Section~\ref{stis_id} below).

\begin{figure}
\resizebox{\hsize}{!}{\includegraphics{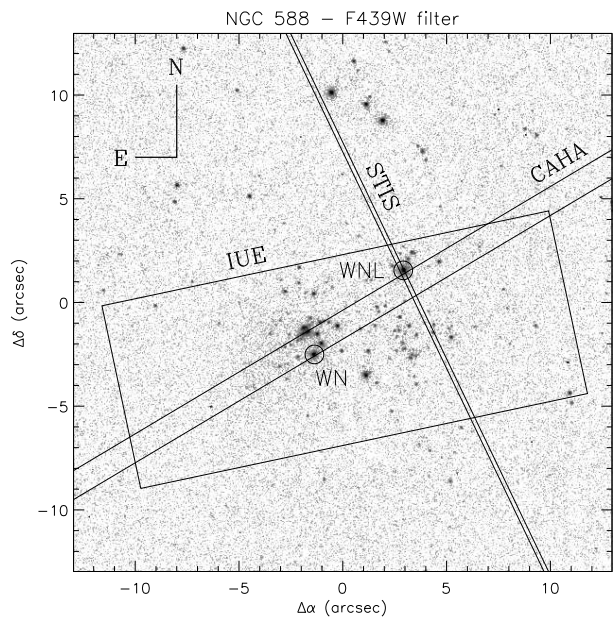}}
\caption{HST image of \object{NGC 588} in filter F439W and the spectral slits.
The peculiar stars (WNL, WN) are discussed in Section~\ref{photappr}.}
\label{n588_b}
\end{figure}

\section{Additional spectroscopic and photometric processing}
\label{datproc}

\subsection{Nebular extinction}
Once the spectra were reduced, we performed a self-consistent procedure to
deredden the nebular lines, based on the Balmer lines H$\alpha$ through
H$\delta$. First, the emission line fluxes were corrected for foreground
Galactic extinction, which amounts to $E_{B-V}=0.045$ \citep{bh84}, the
Galactic law being here the ensemble of the \cite{s79} ultraviolet law and of
the optical law of \cite{nth75} (as shown in Seaton's paper). Then a reddening
law was fitted to the thus corrected Balmer decrement, taking into account the
photometric and measurement errors of these emission lines\footnote{A value of
the electron temperature of 11000 K \citep{vpd88} was assumed for the
theoretical Balmer ratio.}. This process was performed for each spatial
increment along the slit independently, thus obtaining the reddening along the
slit. The residuals from the actual line ratios to the fitted law were then
plotted along the slit, giving statistical fluctuations around a mean offset
value (due to a small residual photometric calibration) except at those
locations where the underlying stellar population has a significant
contribution of Balmer lines in absorption. The equivalent width of the lines
at these locations was thus computed and the correction applied to the
intensity of the emission lines, then computing the corrected value of
reddening. Assuming the LMC extinction law \citep{h83}, a priori adequate due
to the metallicity of the object \citep{vpd88}, we derived an $E_{B-V}$ curve
along the slit, and found it to be almost constant, with a value of
$0.11\pm0.02$ around the stellar cluster. This result is virtually independent
of the selected extinction law, since different laws predict nearly the same
relative extinctions within the H$\delta$--H$\alpha$ wavelength range.

\subsection{Spatial registration between spectra and images}
\label{spareg}

We have registered precisely the position of the CAHA slit with respect to the
HST images. This is achieved by convolving the HST images with the atmospheric
point spread function (PSF) at the time of the ground based observations, and
comparing the spatial flux distributions. In order to compute the atmospheric
PSF we modeled the spatial flux distribution of a point source in the 2D
spectral frames in the following way. We assumed the atmospheric PSF to be a
\cite{m69} function\footnote{This has been shown to be the best analytical
representation of atmospheric turbulence \cite[e.g., ][]{t91}},
$h(r)=(1+(r/R_c)^2)^{-\beta}$. We selected a standard star observed just before
\object{NGC 588}, and examined its profile along the slit around 4400 {\AA}.
This profile was clearly asymmetrical, indicating image distortion inside the
spectrograph. We found that the observed star profile is very well reproduced
by a model of the form $H(x)=\sum_{i=1}^{3}a_iM(x-c_i)$, where $x$ is the
spatial increment variable, $c_i$ the spatial location of the peak of each
component and $a_i$ its weight, and $M(x)$ is the Moffat PSF integrated across
the slit width. Thus, we assumed that the signal of the observed object was
convolved by a Moffat function determined by turbulence, truncated by the slit,
and split into three components within the spectrograph, before hitting the
detector. The same analysis near 6600 {\AA} and 8700 {\AA} led to similar
models. Once the PSF model was established, we proceeded as follows.

The blue 2D spectral frame was multiplied by the response of the HST filter
F439W and the result integrated along the wavelength axis; in this way we
obtained from the spectral frames the spatial flux distribution of \object{NGC
588} through the HST filter F439W. Then, we applied our image degradation model
to the HST image for different values of the Moffat parameters $R_c$ and
$\beta$ (that tend to vary with time), placed a synthetic slit on the blurred
image, and extracted the profile integrated across the slit. $R_c$, $\beta$ and
the central position of the slit were fit by maximizing the correlation between
the profiles extracted from the spectrum and from the image. We found
$R_c\approx 1.1\arcsec$ and $\beta\approx 1.5$, corresponding to a seeing of
FWHM$\sim1.6$ arcsec. The PSF and the slit position were used to calculate the
aperture throughputs used in Section~\ref{classi} and \ref{photappr}.

We also used the two-dimensional frames and the F170W HST/WFPC2 image to
determine accurately the slit position and angular resolution of the IUE data,
assuming a Gaussian PSF and using the same technique as for the optical
spectra. The resulting correlation is very good for a FWHM of 6.4$\arcsec$,
with however a strong photometric mismatch between the data: the IUE flux
integrated within the F170W band was found to be higher than the HST one by a
factor 1.41. This mismatch between IUE and HST absolute calibrations has been
found also for \object{NGC 604} \citep{gdp00}. HST absolute photometry is
accurate to a few \%, so we decided not to correct the HST image, but to divide
the IUE spectra by the factor 1.41.

\subsection{Nebular continuum subtraction}
\label{nebsub}
To analyze the spectral properties of the cluster itself, we removed the
nebular continuum from our spectra. For this, we calculated it according to
\cite{o74}, including the HI free-free, free-bound and two-photon components as
well as the HeI free-free and free-bound emissions, assuming a gas temperature
of 11000 K \citep{vpd88}; we reddened the computed nebular continuum to match
the observations, normalized it with the nebular H$\beta$ emission line
intensity in the spatial zone of the optical slit selected to study the cluster
(knowing the H$\beta$ equivalent width with respect to the nebular continuum),
and removed it.

\subsection{Aperture throughput correction}
\label{apcorr}
Once the optical and UV slit positions and angular PSFs were determined (see
Section \ref{spareg}), we calculated the aperture throughputs of the slits for
the whole cluster, i.e. the fraction of light transmitted by these slits. For
this, we computed the sky background of each image by averaging, at a given
position, the pixels contained in a rectangular box and close enough to the
median value of this box. The resulting map of the sky and of the nebula was
removed from the original in order to keep only the stellar signal. The
remaining stellar images were then convolved by the optical or UV PSFs, and in
each band, we compared the total flux of the cluster to the one entering the
slit. In optical, the aperture throughputs were found to be independent of the
band (F170W, F336W, F439W or F547M), and we retained a mean value of 0.169 for
a slit width of 1.2'', valid in the whole optical range. This throughput was
proportional to the chosen slit width, known to $\approx$20\% accuracy. The
uncertainty on the PSF led to an uncertainty of 10\% on the throughput. For the
IUE spectra, we found a throughput of 0.83 in the F170W filter. This means that
we had to divide the CAHA spectra by 0.169, and the IUE spectra, by 0.83, in
order to obtain the integrated spectrum of the whole cluster. We also
calculated the aperture throughputs for nebular emission: 0.062 for the optical
spectra, and 0.44 for IUE data.

\section{Integrated canonical modeling of the cluster}
\label{classi}
The cluster was first modeled in the classical way, i.e. as a stellar
population whose IMF is rigorously a continuous power law $dN/dm\propto
m^{-\alpha}$, $m_{\rm low}\leq m\leq m_{\rm up}$. Assuming instantaneous star
formation, we used Starburst99 \citep{lsg99}, with $m_{\rm low}=1$ M$_{\sun}$,
making use of standard mass-loss evolutionary tracks at metallicities Z=0.004
\citep{cmm93} and Z=0.008 \citep{smm93}, close to the metallicity of the gas
\citep{vpd88} (the solar value is here Z=0.02). We selected the atmosphere
models of \cite{lbc97}, \cite{hm98} and \cite{phl01}, the latter two ones
having been compiled by \cite{snc02}. The age $\tau$ of the cluster, its
metallicity Z, and the IMF slope $\alpha$ and upper limit $m_{\rm up}$ were
constrained with observables available for the cluster as a whole. These are:
i) the UV and optical colors, integrated over the entire cluster; ii) the
H$\beta$ equivalent width $W({\rm H}\beta)$ measured over the whole object
(nebula and cluster); iii) the profiles of the UV stellar lines present in the
IUE spectrum; iv) the presence or absence of the main Wolf-Rayet (WR) emission
lines in the optical SED.

\subsection{SED colors and derived extinction}
\label{sedcol}
The SED colors were used to determine the amount and the law of extinction of
the stellar flux. Indeed, the color excess $E_{B-V}$ derived from the Balmer
lines of a nebula has been found not to apply systematically to its ionizing
source \cite[e.g., ][]{mhk99}. Furthermore, whereas different extinction laws
generally do not differ significantly from each other in the optical range and
cannot be discriminated by the use of the nebular Balmer lines, they strongly
vary in the UV counterpart. Fortunately, the intrinsic UV and optical colors of
an OB association little depend on the age and IMF of the latter, and can be
used to characterize the extinction, provided that one knows approximately the
main characteristics of this association. \cite{mhk99} found an age of 2.8 Myr
for a very flat IMF ($\alpha=1$) with an upper mass cutoff of 120 M$_{\sun}$,
for the cluster of \object{NGC 588}. We computed SEDs for a range of ages, IMF
slopes and upper mass cutoffs around these values, and calculated the
corresponding fluxes in the four bands F547M, F439W, F336W and F170W. Then, the
colors of the cluster were measured on the HST images cleared of their diffuse
background. Table~\ref{tab_col} shows the measured colors and some model
outputs. The differences of color between the observations and the models were
attributed to interstellar reddening. We tested three extinction laws: Galactic
\citep{s79,nth75}, LMC \citep{h83} and SMC \citep{plm84,blm85}, whose values at
the barycentric wavelengths of the bands are summarized in Table~\ref{tab_col}.
For any set of model colors, the best fit was obtained with the SMC law and
$E_{B-V}=0.08\pm0.04$ (the error bar includes the uncertainties in the
observational data and the spread in the theoretical colors). This value is
compatible with the amount of extinction derived from the nebular Balmer lines,
meaning that there is no indication of differential extinction between the
stellar and the nebular fluxes. Since the nebular value, $E_{B-V}=0.11\pm0.02$,
was determined more accurately, we have also adopted it for the cluster in the
following. Considering these results and the 800 kpc distance of \object{M33}
\citep{lsg02}, we derived the cluster luminosity in the F439W band:
4.71$\times$10$^{35}$ erg\,s$^{-1}$\,{\AA}$^{-1}$.
\begin{table}
\begin{tabular}{ccccc}
Filter & F547M & F439W & F336W & F170W \\
$\lambda$ ({\AA})& 5470 & 4300 & 3330 & 1740 \\
\hline
Obs. $M_{547}-M_\lambda$ & 0.00 & 0.84 & 1.48 & 3.03 \\
 & $\pm$0.03 & $\pm$0.03 & $\pm$0.04 & $\pm$0.05 \\
\hline
$M_{547}-M_\lambda$ (a) & 0.00 & 0.96 & 1.85 & 4.10 \\
$M_{547}-M_\lambda$ (b) & 0.00 & 0.87 & 1.49 & 3.22 \\
$M_{547}-M_\lambda$ (c) & 0.00 & 0.82 & 1.19 & 2.94 \\
\hline
$f(\lambda)$ (Galactic) & 3.21 & 4.18 & 5.06 & 7.80 \\
$f(\lambda)$ (LMC) & 3.10 & 4.13 & 5.37 & 9.25 \\
$f(\lambda)$ (SMC) & 2.72 & 3.65 & 4.65 & 10.51 \\
\end{tabular}
\caption{Observed and model colors with filter F547M as the zero point. The
observed colors are here corrected for foreground Galactic extinction. The
three models (a), (b) and (c) shown here stand for $m_{\rm up}=120$ M$_{\sun}$,
Z=0.004, and respectively, $\alpha=1$, 1 and 2.35 and $\tau=$2, 3 and 4 Myr. We
also show the three tested extinction laws in the form
$f(\lambda)=A(\lambda)/E_{B-V}$. The color differences $\Delta(M_y-M_\lambda)$
between the observations and a given model were to be well fit by a function of
the form $a+E_{B-V}f(\lambda)$ for the tested extinction law to be accepted.}
\label{tab_col}
\end{table}

\subsection{H$\beta$ equivalent width}
This quantity is often used in modeling of massive stellar populations, as a
way to characterize the ratio of the ionizing photon rate to the optical
continuum. It is mainly sensitive to the global distribution of effective
temperatures of the stars, although it also depends on the fraction of the
ionizing flux absorbed by the nebular gas. Since this fraction is a priori
unknown, we decided only to require the model to predict $W({\rm H}\beta)$
greater than or equal to the observed value.

{\bf The value of $W({\rm H}\beta)$ used here is the ratio of the total nebular
H$\beta$ flux to the stellar H$\beta$ continuum integrated over the cluster.
The integrated values of these two quantities were obtained by dividing the
values measured in the blue CAHA spectrum (cleared of the nebular continuum;
see Section~\ref{nebsub}) by the aperture throughputs of the optical slit for
the nebular gas and for the stellar continuum (Section~\ref{apcorr}). This
estimate of $W({\rm H}\beta)$ is affected by the error in the
nebular-to-stellar ratio of the two aperture throughputs, evaluated to
$\approx$10\%. We found $W({\rm H}\beta)=330\pm30$~{\AA}.}

\subsection{UV lines}
The UV lines that form in the atmospheres of massive stars were studied by
\cite{sa87a}. They showed the correlation existing between the shapes and
intensities of the Si{\sc IV} $\lambda$1400 and C{\sc IV} $\lambda$1550 lines
and the spectral type and luminosity class of early-type stars. The use of
these two lines and, in some cases, the He{\sc II} $\lambda$1640 line, can be a
good complement to $W({\rm H}\beta)$, for the determination of the evolutionary
status and IMF of the cluster, due to their sensitivity to the luminosity class
of the stars \citep[e.g., ][]{sa87b,rlh93,mhk91,mhk99}.

The IUE short-wavelength spectrum of \object{NGC 588} has low resolution and
signal-to-noise ratio, and was not used directly in the fit of the population
parameters. Instead, it served to discuss and possibly reject some models
determined with the other observables, by visual comparison between model and
observed spectra. Furthermore, we assumed this spectrum to be representative of
the whole cluster, since the IUE slit covers nearly all the flux of the cluster
(see Section \ref{apcorr}).

\subsection{WR features}
The optical spectra clearly show the broad He{\sc II} $\lambda4686$ emission
line produced by one or more WR stars. No corresponding C{\sc IV} $\lambda5812$
feature was found, showing that the detected WR stars are of WN type. The
presence of such stars in the cluster fixes its age to 3.2--4.4 Myr if Z=0.004,
and 2.8--4.5 Myr if Z=0.008, according to Starburst99. These age ranges are
valid for $m_{\rm up}=120$ M$_{\sun}$, and are narrower for lower mass cutoffs.

Since the optical slit covers only a fraction of the cluster, our optical
spectra do not necessarily include the signatures of all the  kinds of stars
present in the cluster. Consequently, we were unable to estimate the total
number of WN stars in the entire cluster. Therefore, to constrain the models,
we only used the fact that this number is of at least 1. Likewise, we ignored
the non-detection of WC star in the optical spectrum, and did not process the
number of WC stars predicted by the models.

\subsection{Model fitting and results}
\label{res_class}
\begin{figure}
\resizebox{\hsize}{!}{\includegraphics[angle=90]{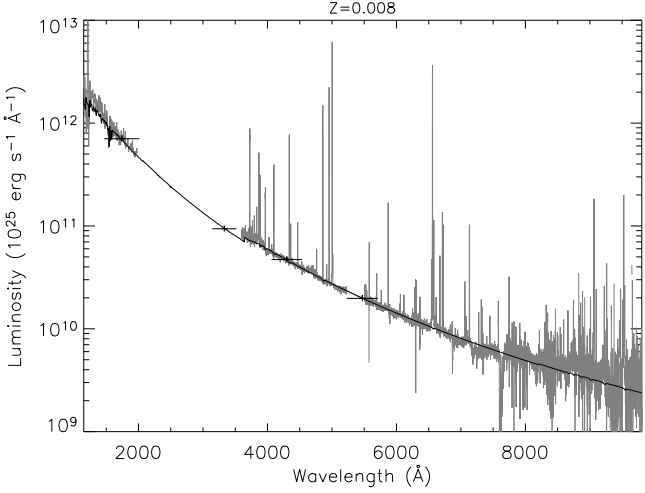}}
\caption{Observed SED vs. SED constructed with the best-fit parameters. {\bf
The observed SED is here supposed to be representative of the whole cluster,
even though the slits (in particular the optical one) only sample a portion of
the cluster.}}
\label{an_sed}
\end{figure}
We performed a chi-square-like fit of $\tau$, $\alpha$ and Z over $W({\rm
H}\beta)$ and the number of WN stars ${\mathcal N}({\rm WN})$, using the
outputs of the Starburst99 code. ${\mathcal N}({\rm WN})$, the total mass
$M_{\rm tot}$ and the number ${\mathcal N}({\rm O})$ of O stars were multiplied
by the coefficient needed to reproduce the luminosity of the cluster in filter
F439W. The chi-square formula we used is the following:
\begin{eqnarray}
\chi^2=\left(\frac{\max(330-W({\rm H}\beta),0)}{30}\right)^2\nonumber\\
-2\,\ln\left(1-e^{-{\mathcal N}({\rm WN})}\right)
\end{eqnarray}
The second term of the right part of this equations comes from the fact that
the models described in this section predict numbers of WN stars, ${\mathcal
N}({\rm WN})$, that are not integers. Then, we treated ${\mathcal N}({\rm WN})$
as the expectancy of a Poisson statistic process, and considered the likelihood
to obtain at least one WN star with such a process, which is $P_{\rm
WN}=1-\exp(-{\mathcal N}({\rm WN}))$.
\begin{table}
\begin{tabular}{cccc}
& Z=0.004 & Observed & Z=0.008 \\
\hline
$W({\rm H}\beta)$ ({\AA}) & 204 & 330$\pm$30 & 247 \\
${\mathcal N}({\rm WN})$ & 0.17 & $\geq$1 & 0.03 \\
reduced $\chi^2$ & 10 & & 7 \\
\hline
$\tau$ (Myr) & 3.5 (3.4--3.6) & & 2.8 (2.8--3.0) \\
$\alpha$ & 2.35 (2--3.1)& & 2.35 (2--3) \\
$m_{\rm up}$ (M$_{\sun}$) & 110 ($\geq$90) & & 120 ($\geq$110) \\
\hline
$M_{\rm tot}$ (M$_{\sun}$) & ($\approx$3000) & & ($\approx$3000) \\
${\mathcal N}({\rm O})$ & ($\approx$10) & & ($\approx$12)
\end{tabular}
\caption{Results of the canonical analytical modeling of the cluster, for the
two metallicities Z=0.004 and Z=0.008. Single values correspond to the best
fits. Parenthesized ranges correspond to the approximate 3$\sigma$ ranges.}
\label{tab_anfit}
\end{table}

\begin{figure}
\resizebox{\hsize}{!}{\includegraphics{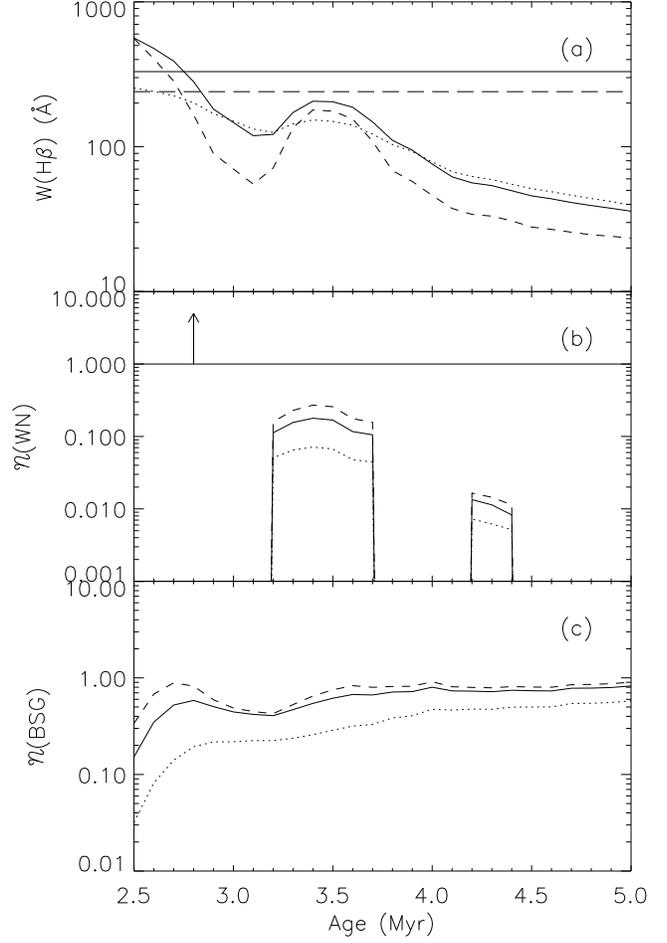}}
\caption{Evolution of $W({\rm H}\beta)$ (top), ${\mathcal N}({\rm WN})$
(middle) and ${\mathcal N}({\rm BSG})$ (bottom) with the cluster age, for
Z=0.004, $m_{\rm up}=120$ M$_{\sun}$ and the IMF slopes $\alpha=$2.35 (full
lines), 1 (short-dashed lines) and 3.3 (dotted lines). In the $W({\rm H}\beta)$
panel, the full horizontal line shows the observed value, and the long-dashed
one, the 3$\sigma$ limit. In the ${\mathcal N}({\rm WN})$ panel, the horizontal
line and the arrow are here to remind that at least one WN star is present in
the cluster.}
\label{fig_whbwnbsg004}
\end{figure}

\begin{figure}
\resizebox{\hsize}{!}{\includegraphics{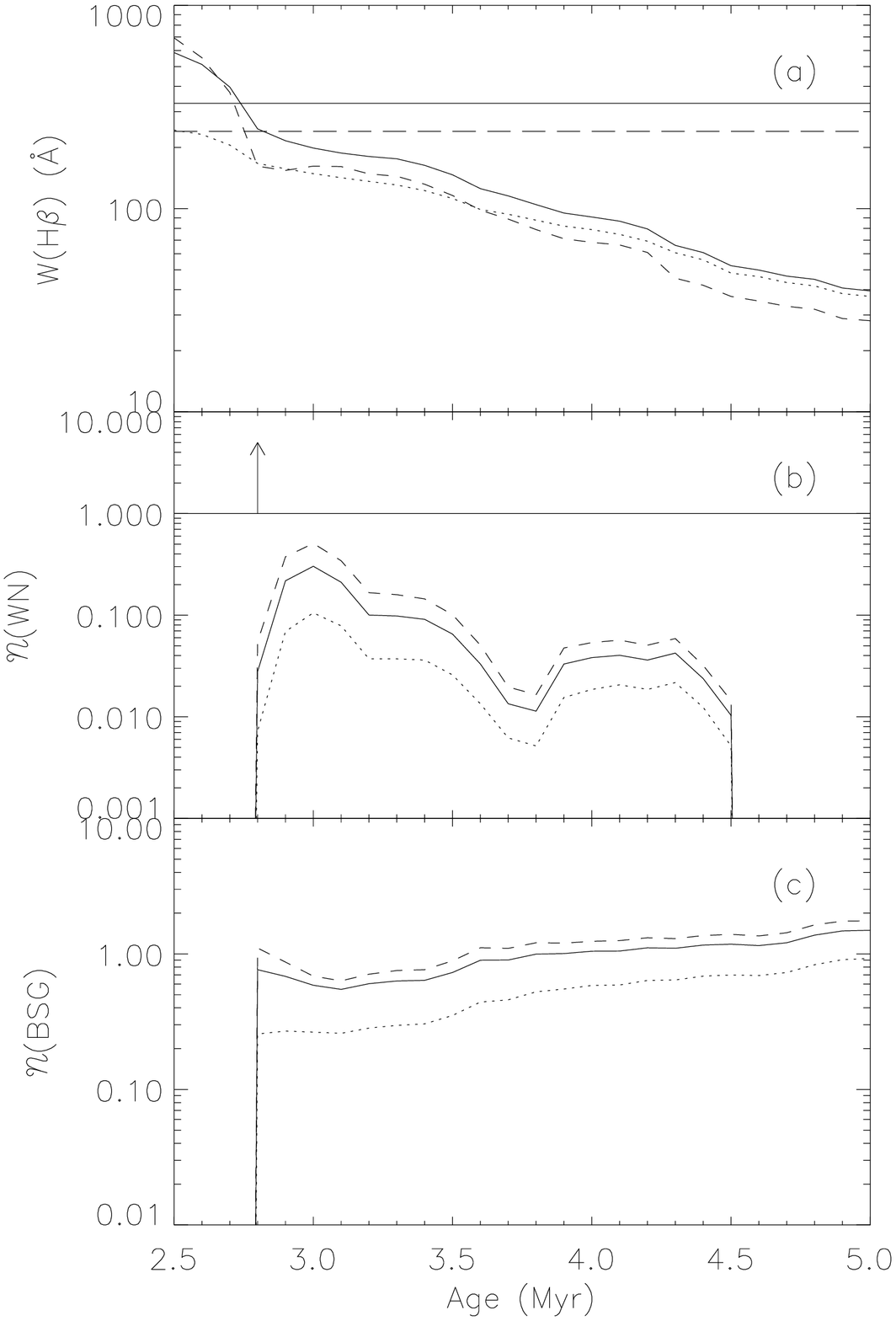}}
\caption{Same as Fig. \ref{fig_whbwnbsg004} for Z=0.008}
\label{fig_whbwnbsg008}
\end{figure}

The results of the fit are summarized in Table~\ref{tab_anfit}. The model
associated to the minimum $\chi^2$ is described by $\tau=$2.8 Myr, $m_{\rm
up}=$120 M$_{\sun}$ and Z=0.008. However, this model, with a reduced $\chi^2$
as high as 7, shows significant discrepancies with the observations: both
${\mathcal N}({\rm WN})$ and $W({\rm H}\beta)$ are well below the observed
value. We can also note that due to the small value of ${\mathcal N}({\rm
WN})$, the model He{\sc II} $\lambda$4686 WR bump is negligible (see Fig.
\ref{an_sed}). The bad fit of $W({\rm H}\beta)$ when ${\mathcal N}({\rm WN})$
is high enough can be attributed to the presence of blue supergiants (BSGs),
defined here as having effective temperatures $T_{\rm eff}\leq 3\times 10^4$ K,
in the model cluster such that each of them, being alone in the nebula, would
cause $W({\rm H}\beta)\lesssim 100$ {\AA}: according to the best-fit model,
their number is ${\mathcal N}({\rm BSG})=0.8=0.07\,{\mathcal N}({\rm O})$. In a
general way, such stars were predicted at any age with high enough values of
${\mathcal N}({\rm WN})$, resulting in the absence of a good compromise between
$W({\rm H}\beta)$ and ${\mathcal N}({\rm WN})$; this is illustrated by Fig.
\ref{fig_whbwnbsg004} and \ref{fig_whbwnbsg008}.


\begin{figure}
\resizebox{\hsize}{!}{
\includegraphics[angle=90]{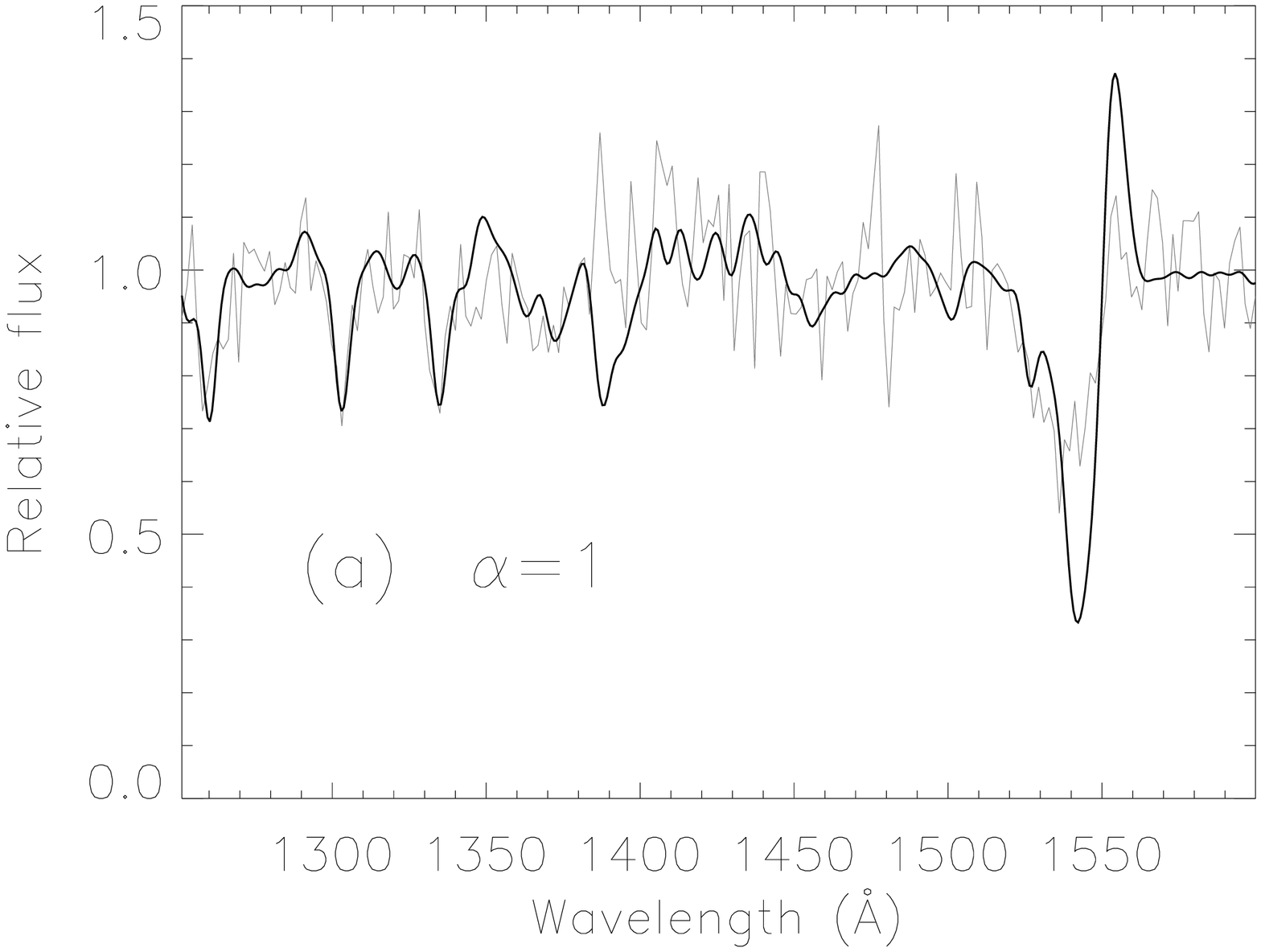}
\includegraphics[angle=90]{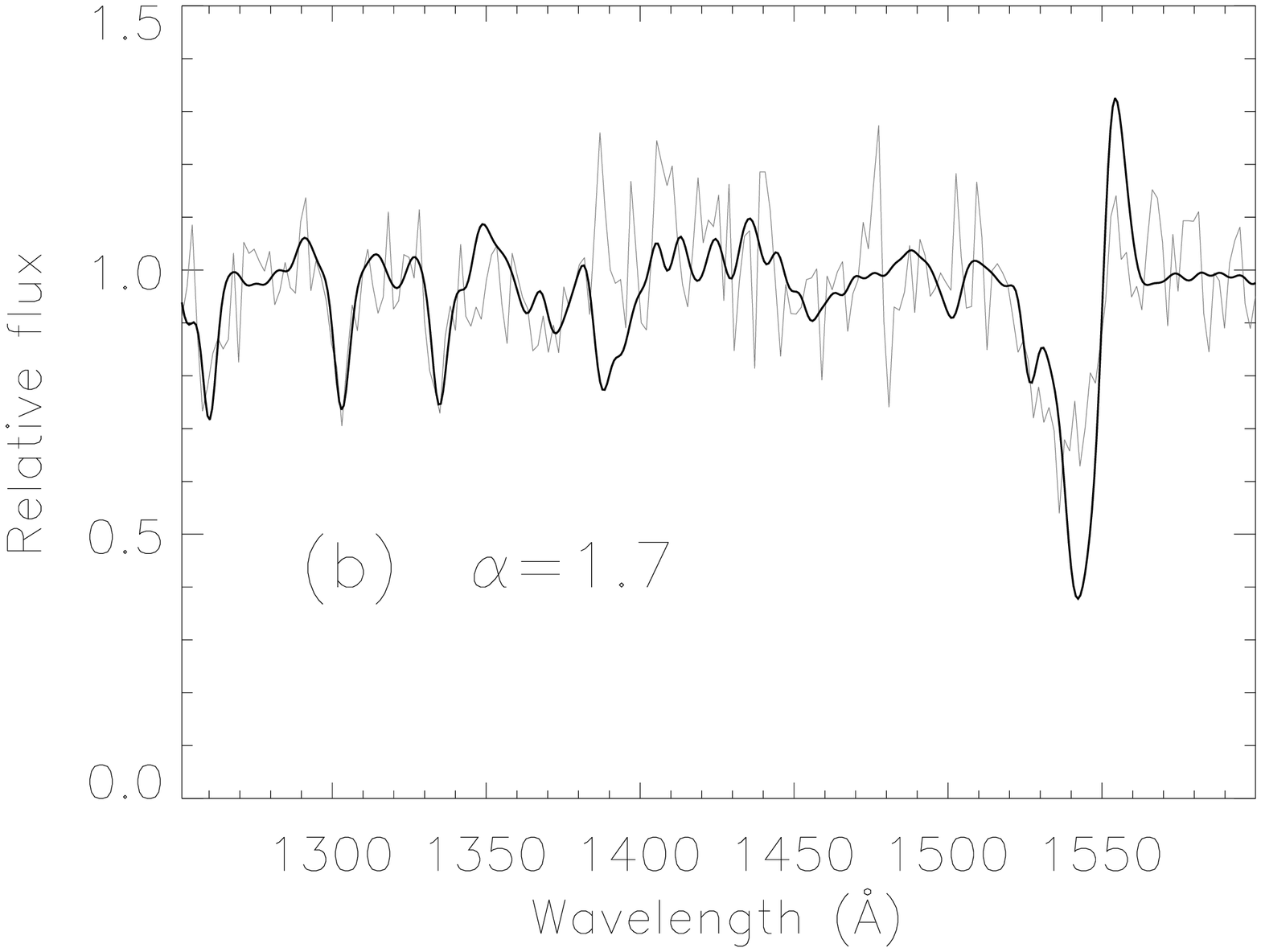}}
\resizebox{\hsize}{!}{
\includegraphics[angle=90]{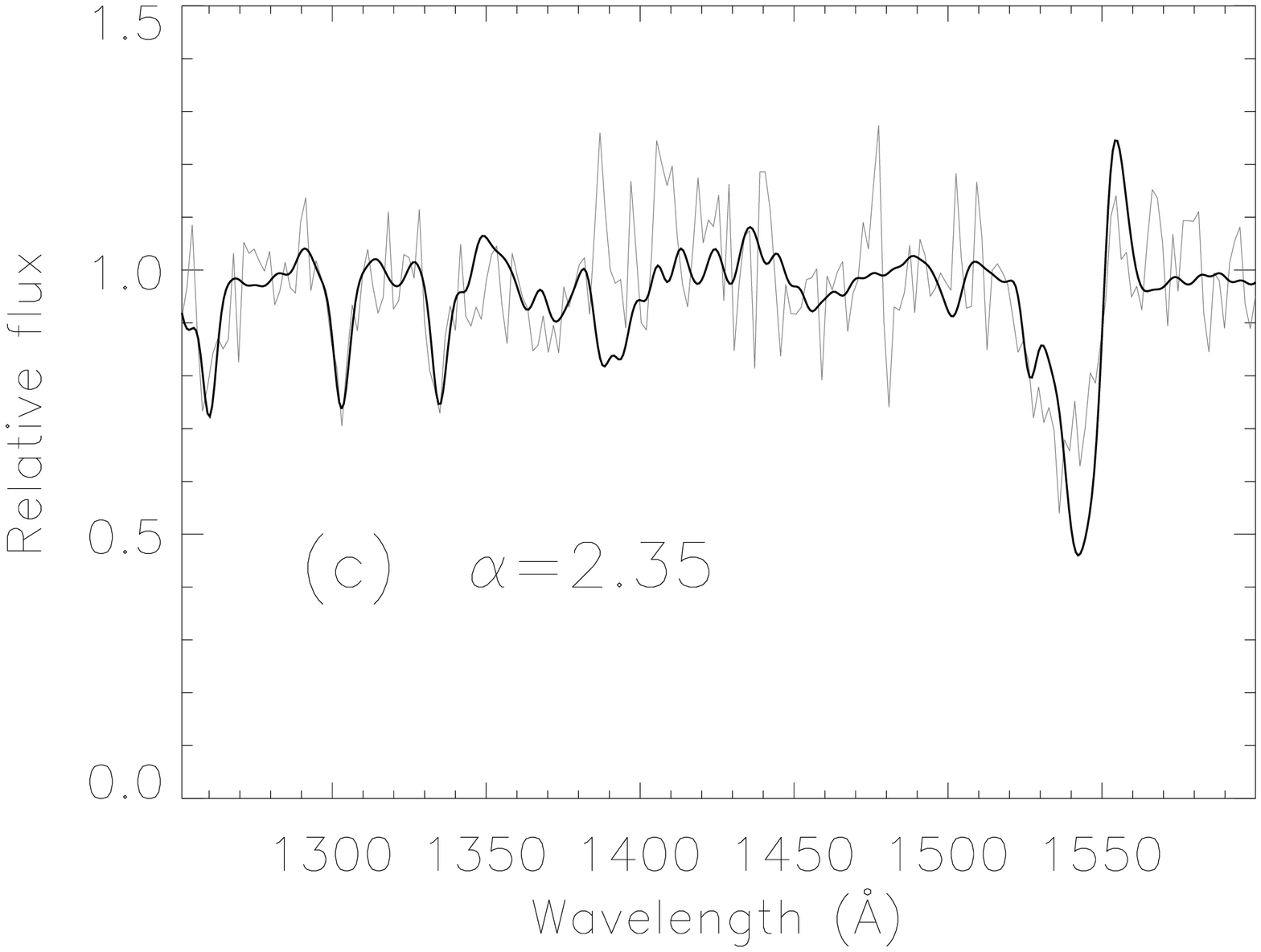}
\includegraphics[angle=90]{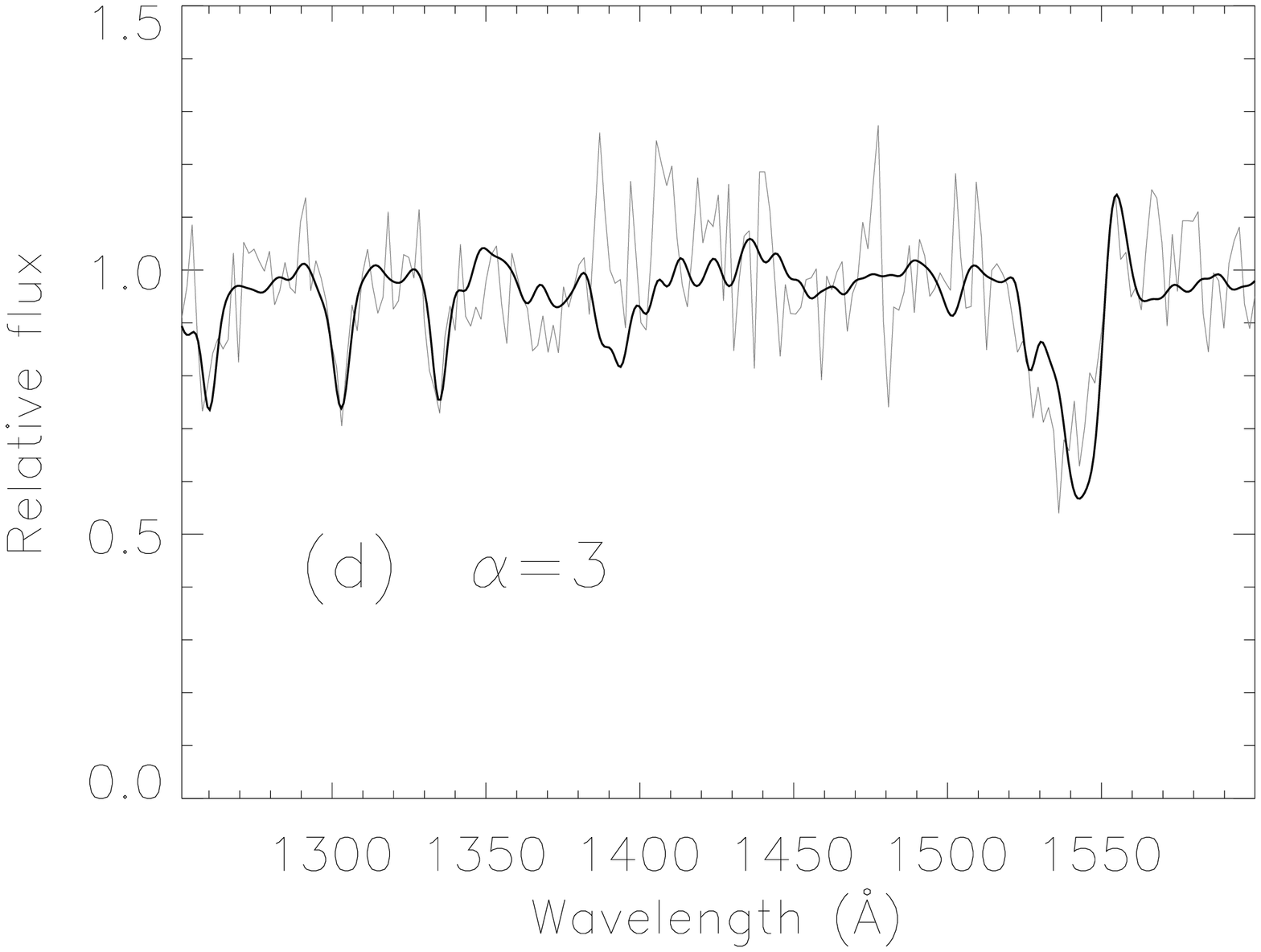}}
\caption{Observed vs. synthetic UV spectra for different IMF slopes. The
synthetic spectra were convolved by a 5 {\AA} FWHM Gaussian curve to match
IUE's resolution.}
\label{an_uvl_alp}
\end{figure}
So far, we were unable to determine the IMF slope more accurately than
constraining it to the range 1.5--3. Now, we show how we can discard IMFs
flatter than $\alpha\approx2$. Fig.~\ref{an_uvl_alp} shows the observed and
model UV spectra in the case of the IMF slopes $\alpha=1$, 1.7, 2.35 and 3, and
for the best-fit metallicity (Z=0.008), age (2.8 Myr) and upper mass limit (120
M$_{\sun}$). The synthetic UV spectra were created with the LMC/SMC library of
Starburst99 \citep{llh01}. The depths of the Si{\sc IV} $\lambda$1400 and C{\sc
IV} $\lambda$1550 lines were found to decrease with increasing IMF slope, due
to the conservation of their equivalent widths and increase of their spectral
widths. This result is in agreement with \cite{rlh93}. {Due to excessive depths
of the lines, the cases $\alpha=1$ and 1.7 were rejected}, while we considered
the fits reasonably good for the other, steeper IMF slopes.

\section{Need to account for statistical fluctuations}
\label{statmod}
In the previous section, we obtained a model attempting to reproduce as well as
possible $W({\rm H}\beta)$, the numbers of WN stars and the UV stellar lines in
the framework of a classical cluster analysis, that only takes into account the
accuracy of the observational data. According to this model, the mass of the
cluster is $\approx$3000 M$_{\sun}$. Meanwhile, assuming a metallicity Z=0.004
and using the same evolutionary tracks, \cite{mhk99} obtained a cluster age
$\tau=2.8$, an IMF slope $\alpha=1$ and a total mass of 534 M$_{\sun}$, by the
analysis of, mainly, the IUE-SWP and IUE-LWR spectra, but in absence of
knowledge about the WR content of the cluster.

In both works, the results have been obtained \emph{ under the following
assumptions}: correct stellar evolutionary tracks, instantaneous burst
hypothesis and the implicit existence of a zero age main sequence \citep[but
see][]{tps03} and a number of stars in the cluster large enough to perfectly
sample the IMF. However the estimated masses in both cases are about two orders
of magnitude below the 10$^5$ M$_{\sun}$ mass for which, according to
\cite{cmh94}, the whole IMF would be well filled. Hence, it is expected that
sampling effects in the IMF will strongly affect any classical integrated
analysis of this cluster.

Furthermore, the F439W luminosity of the cluster, with a value of -10.08 mag.
is lower than the luminosity of the brightest star in the isochrone of the
fitted model, -11.42 mag., that \cite{cl04} showed to be the lowest luminosity
limit under which a cluster cannot be modeled with classical population
synthesis. Even worse, this limit is brighter than the entire cluster by as
much as 1.4 magnitude. This implies that the results of classical synthesis
models regarding magnitudes and colors can be strongly biased \citep{cv02}.
Unfortunately, the actual theoretical status of statistical modeling of stellar
clusters is not ready yet to solve such a situation. The only realistic
assessments that can be robustly obtained are: (i) due to the presence of at
least one WN star, the age range is between 2.8 and 4.5 Myr (assuming there is
no issue in isochrone computations and that WR star formation is not due to the
evolution of binary systems), and (ii) the amount of gas transformed into stars
at the onset of the burst does not exceed about 10$^4$ M$_{\sun}$ (for a
Salpeter IMF and $m_{\rm low}=1$ M$_{\sun}$), for which the luminosity of the
cluster would reach the lowest luminosity limit.

This implies that the discussion about ages, IMF and discrepancies between the
models and the observations in the analysis made with standard synthesis models
hardly makes sense in our case.

One of our main objectives is to infer a plausible shape of the ionizing
continuum, which implies the use of a technique different from the standard
modeling as we show in the next section.

\section{Star-by-star characterization of the cluster}
\label{photappr}
Due to the small number of stars influencing the observables and the ionizing
flux of the \object{NGC 588} cluster, and because of the unsatisfactory results
of the analysis detailed in Section~\ref{classi}, we decided to study not only
the cluster as a whole, but also its individual stars. We achieved photometry
of these stars, used isochrone curves to parameterize the cluster, attributed a
model to each star, and synthesized spectra to be compared to the observations.

\subsection{Stellar photometry}
\label{stelphot}
We performed photometric measurements of the stars on the five HST images,
using the \texttt{noao.digiphot.daophot} package of IRAF. In each band, we ran
\texttt{daofind} and \texttt{phot} to detect candidate stars and measure their
intensities in 6 pixel radius apertures. The brightest isolated stars served to
compute analytical PSFs for all the bands. We then ran \texttt{allstar} to
reject the artifacts of the \texttt{daofind} routine and measure more
accurately the intensities of the remaining objects. We selected again bright
and isolated stars to carefully calculate the aperture corrections for the
bands with the \texttt{mkapfile} routine of the \texttt{noao.digiphot.photcal}
package, and derive the stellar intensities integrated over the full PSF
extents. The intensities were corrected for charge-transfer inefficiency and
filter contamination (see \S~\ref{hstimg}), and converted into apparent
magnitudes with the DN-to-flux keyword PHOTFLAM of the image headers,. The
apparent magnitudes were converted into absolute (but reddened) magnitudes with
the distance modulus 24.52 \citep{lsg02}. The densest part of the cluster was
not resolved by \texttt{daofind}, but from visual inspection, we found it to
contain five clearly distinct stars. Using the Levenberg-Marquardt's
least-square method, and already knowing the PSF mathematical models, we fitted
the positions and fluxes of these five stars in each of the images; we detected
no other star in the residuals of the fits, and retained our five-star
detection. The final photometric lists were composed of 536 stars in band
F547M, 698 in F439W, 353 in F336W, 574 in F170W and 179 in F469N. 173 of them
were common to bands F547M and F439W, 56, to F547M, F439W, F336W and F170W, and
20 to all bands. Table~\ref{tab_phot} lists the 56 stars common to F547M,
F439W, F336W and F170W. In terms of absolute magnitudes, the 3$\sigma$
detection limits were estimated to be approximately $0.8$, $-0.1$, $-0.8$,
$-3.8$ and $-2.8$ in filters F547M, F439W, F336W, F170W and F469N,
respectively.

\begin{table*}
\begin{tabular}{cccccccccc}
Star \# & R.A. & Dec. & $M_{547}$ & $M_{439}$ & $M_{336}$ & $M_{170}$ &
$M_{469}$ & $E_{B-V}$ & $m_{\rm ini}$\\
 & 01h32m & +30$\degr$ & & & & & & &(M$_{\sun}$)\\
\hline
 1&45.23s&38$'$58.4$''$&$-6.63\pm$0.04&$-7.42\pm$0.03&$-8.14\pm$0.04&$-9.34\pm$0.04&$-7.26\pm$0.06&$0.11\pm$0.03$^{~}$&55.3\\
 2&45.56s&38$'$54.4$''$&$-6.03\pm$0.04&$-6.91\pm$0.04&$-7.74\pm$0.03&$-9.32\pm$0.04&$-6.68\pm$0.08&$0.00\pm$0.03$^{~}$&55.5\\
 3&45.50s&39$'$07.0$''$&$-6.12\pm$0.04&$-6.91\pm$0.03&$-7.55\pm$0.04&$-8.57\pm$0.05&$-7.09\pm$0.06&$0.11\pm$0.02$^{~}$&43.8\\
 4&45.31s&39$'$05.7$''$&$-5.60\pm$0.05&$-6.45\pm$0.04&$-7.08\pm$0.04&$-8.24\pm$0.07&$-6.19\pm$0.09&$0.11\pm$0.03$^{~}$&40.1\\
 5&46.81s&39$'$06.9$''$&$-5.34\pm$0.07&$-6.22\pm$0.05&$-6.89\pm$0.06&$-8.26\pm$0.08&$-5.91\pm$0.14&$0.08\pm$0.04$^{~}$&38.1\\
 6&45.37s&38$'$53.4$''$&$-5.34\pm$0.07&$-6.19\pm$0.07&$-6.89\pm$0.06&$-8.33\pm$0.06&$-5.57\pm$0.13&$0.08\pm$0.04$^{~}$&37.1\\
 7&45.58s&38$'$55.5$''$&$-4.86\pm$0.05&$-5.90\pm$0.05&$-6.55\pm$0.05&$-8.29\pm$0.06&$-5.02\pm$0.08&$0.02\pm$0.04$^{~}$&32.1\\
 8&45.37s&39$'$06.4$''$&$-4.80\pm$0.07&$-5.59\pm$0.06&$-6.45\pm$0.06&$-7.73\pm$0.10&$-5.36\pm$0.15&$0.11\pm$0.02$^a$&33.1\\
 9&45.53s&38$'$54.9$''$&$-4.64\pm$0.07&$-5.49\pm$0.06&$-6.35\pm$0.05&$-7.96\pm$0.07&$-5.04\pm$0.15&$0.07\pm$0.01$^a$&32.1\\
10&45.60s&38$'$55.6$''$&$-4.56\pm$0.06&$-5.31\pm$0.06&$-6.17\pm$0.06&$-7.46\pm$0.09&$-4.40\pm$0.08&$0.12\pm$0.01$^a$&30.1\\
11&45.59s&38$'$55.4$''$&$-4.27\pm$0.07&$-5.26\pm$0.06&$-5.78\pm$0.09&$-7.46\pm$0.09&$-4.85\pm$0.09&$0.09\pm$0.01$^a$&29.1\\
12&45.47s&38$'$55.7$''$&$-4.42\pm$0.06&$-5.19\pm$0.05&$-6.02\pm$0.06&$-7.51\pm$0.08&$-4.94\pm$0.15&$0.10\pm$0.01$^a$&29.1\\
13&45.55s&38$'$55.3$''$&$-4.28\pm$0.07&$-5.03\pm$0.07&$-5.93\pm$0.07&$-7.39\pm$0.09&$-4.77\pm$0.17&$0.10\pm$0.01$^a$&27.1\\
14&45.59s&38$'$55.7$''$&$-4.02\pm$0.09&$-4.90\pm$0.07&$-5.39\pm$0.09&$-6.93\pm$0.13&$-4.79\pm$0.09&$0.12\pm$0.02$^a$&26.1\\
15&46.07s&39$'$02.5$''$&$-4.10\pm$0.06&$-4.79\pm$0.06&$-5.51\pm$0.07&$-6.54\pm$0.14&              &$0.18\pm$0.02$^{~}$&27.1\\
16&45.60s&38$'$55.4$''$&$-3.87\pm$0.08&$-4.75\pm$0.07&$-5.37\pm$0.08&$-6.93\pm$0.11&$-4.45\pm$0.09&$0.10\pm$0.02$^a$&24.4\\
17&45.80s&39$'$02.0$''$&$-3.71\pm$0.07&$-4.50\pm$0.07&$-5.32\pm$0.07&$-6.63\pm$0.13&$-3.87\pm$0.27&$0.11\pm$0.02$^a$&21.9\\
18&44.61s&38$'$52.6$''$&$-3.35\pm$0.11&$-4.28\pm$0.08&$-4.88\pm$0.09&$-6.50\pm$0.12&$-3.04\pm$0.50&$0.08\pm$0.02$^a$&20.1\\
19&45.42s&39$'$08.5$''$&$-3.36\pm$0.08&$-4.22\pm$0.08&$-5.00\pm$0.08&$-6.16\pm$0.17&$-3.33\pm$0.42&$0.12\pm$0.02$^a$&19.7\\
20&45.05s&38$'$55.2$''$&$-3.28\pm$0.08&$-4.19\pm$0.07&$-5.01\pm$0.09&$-6.66\pm$0.11&$-3.60\pm$0.36&$0.05\pm$0.02$^a$&19.5\\
21&45.54s&38$'$55.9$''$&$-3.31\pm$0.11&$-4.14\pm$0.08&$-4.86\pm$0.10&$-6.44\pm$0.14&              &$0.08\pm$0.02$^a$&19.1\\
22&45.20s&38$'$59.3$''$&$-3.19\pm$0.13&$-4.07\pm$0.10&$-4.72\pm$0.14&$-6.08\pm$0.19&              &$0.11\pm$0.03$^a$&18.6\\
23&45.10s&38$'$58.9$''$&$-3.05\pm$0.08&$-3.99\pm$0.08&$-4.71\pm$0.10&$-6.13\pm$0.13&$-3.50\pm$0.35&$0.08\pm$0.02$^a$&18.1\\
24&45.67s&38$'$57.4$''$&$-3.09\pm$0.09&$-3.97\pm$0.09&$-4.55\pm$0.12&$-6.08\pm$0.14&              &$0.09\pm$0.02$^a$&17.9\\
25&45.16s&39$'$04.2$''$&$-3.05\pm$0.11&$-3.87\pm$0.11&$-4.15\pm$0.15&$-5.41\pm$0.20&              &$0.18\pm$0.03$^a$&17.2\\
26&45.24s&38$'$56.2$''$&$-2.94\pm$0.09&$-3.81\pm$0.09&$-4.37\pm$0.12&$-5.94\pm$0.15&              &$0.09\pm$0.02$^a$&16.9\\
27&45.36s&38$'$54.6$''$&$-2.86\pm$0.09&$-3.76\pm$0.08&$-4.42\pm$0.11&$-5.87\pm$0.16&              &$0.09\pm$0.03$^a$&16.6\\
28&45.21s&38$'$55.8$''$&$-2.85\pm$0.11&$-3.72\pm$0.09&$-4.28\pm$0.14&$-5.65\pm$0.21&              &$0.12\pm$0.03$^a$&16.3\\
29&46.08s&39$'$01.7$''$&$-2.93\pm$0.10&$-3.71\pm$0.10&$-4.03\pm$0.13&$-4.71\pm$0.32&              &$0.25\pm$0.04$^{~}$&20.0\\
30&45.20s&38$'$54.2$''$&$-3.07\pm$0.18&$-3.62\pm$0.10&$-4.12\pm$0.14&$-5.74\pm$0.19&              &$0.13\pm$0.04$^a$&15.6\\
31&45.62s&38$'$58.6$''$&$-2.75\pm$0.10&$-3.57\pm$0.09&$-4.47\pm$0.11&$-5.91\pm$0.20&              &$0.06\pm$0.03$^a$&15.5\\
32&45.46s&38$'$54.5$''$&$-2.64\pm$0.10&$-3.51\pm$0.09&$-4.26\pm$0.12&$-5.69\pm$0.19&              &$0.08\pm$0.03$^a$&15.1\\
33&45.85s&39$'$07.1$''$&$-2.52\pm$0.11&$-3.34\pm$0.10&$-4.07\pm$0.12&$-5.36\pm$0.24&              &$0.10\pm$0.04$^a$&14.2\\
34&45.33s&38$'$56.7$''$&$-2.39\pm$0.10&$-3.33\pm$0.10&$-4.10\pm$0.12&$-5.73\pm$0.17&              &$0.03\pm$0.03$^a$&14.3\\
35&45.80s&39$'$11.5$''$&$-2.34\pm$0.11&$-3.30\pm$0.09&$-4.01\pm$0.13&$-5.67\pm$0.19&              &$0.03\pm$0.03$^a$&14.1\\
36&45.72s&39$'$10.6$''$&$-2.33\pm$0.13&$-3.27\pm$0.11&$-3.89\pm$0.13&$-5.34\pm$0.23&              &$0.08\pm$0.04$^a$&13.9\\
37&45.51s&38$'$55.9$''$&$-2.42\pm$0.10&$-3.22\pm$0.09&$-3.70\pm$0.14&$-5.26\pm$0.24&              &$0.10\pm$0.04$^a$&13.6\\
38&46.15s&38$'$56.7$''$&$-2.32\pm$0.10&$-3.14\pm$0.11&$-3.74\pm$0.13&$-4.85\pm$0.28&              &$0.14\pm$0.04$^a$&13.2\\
39&45.22s&38$'$54.7$''$&$-2.39\pm$0.16&$-3.09\pm$0.14&$-3.77\pm$0.16&$-4.89\pm$0.29&              &$0.14\pm$0.05$^a$&12.9\\
40&45.12s&38$'$55.8$''$&$-2.40\pm$0.13&$-3.04\pm$0.12&$-3.84\pm$0.15&$-5.12\pm$0.24&              &$0.11\pm$0.04$^a$&12.7\\
41&45.62s&38$'$54.2$''$&$-2.27\pm$0.11&$-3.00\pm$0.11&$-3.69\pm$0.13&$-5.07\pm$0.25&              &$0.10\pm$0.04$^a$&12.6\\
42&45.62s&38$'$57.8$''$&$-2.06\pm$0.13&$-2.99\pm$0.12&$-3.67\pm$0.17&$-5.17\pm$0.25&              &$0.05\pm$0.04$^a$&12.6\\
43&45.61s&38$'$55.1$''$&$-2.39\pm$0.19&$-2.97\pm$0.17&$-3.60\pm$0.19&$-5.06\pm$0.30&              &$0.11\pm$0.05$^a$&12.4\\
44&45.33s&38$'$55.8$''$&$-2.16\pm$0.13&$-2.95\pm$0.12&$-3.70\pm$0.13&$-5.37\pm$0.26&              &$0.03\pm$0.04$^a$&12.4\\
45&44.74s&38$'$55.8$''$&$-1.78\pm$0.15&$-2.94\pm$0.13&$-3.07\pm$0.20&$-5.11\pm$0.29&              &$0.01\pm$0.05$^a$&12.4\\
46&45.33s&38$'$53.0$''$&$-2.23\pm$0.12&$-2.91\pm$0.12&$-3.66\pm$0.16&$-4.92\pm$0.28&              &$0.11\pm$0.05$^a$&12.1\\
47&45.27s&38$'$58.6$''$&$-1.98\pm$0.14&$-2.89\pm$0.13&$-3.56\pm$0.17&$-5.11\pm$0.25&              &$0.04\pm$0.04$^a$&12.1\\
48&45.58s&38$'$55.8$''$&$-2.32\pm$0.20&$-2.88\pm$0.19&$-3.94\pm$0.13&$-5.25\pm$0.26&              &$0.07\pm$0.05$^a$&12.1\\
49&45.01s&38$'$50.9$''$&$-1.90\pm$0.15&$-2.84\pm$0.12&$-3.54\pm$0.18&$-5.34\pm$0.28&              &$0.00\pm$0.05$^a$&12.0\\
50&45.39s&38$'$55.6$''$&$-1.91\pm$0.15&$-2.83\pm$0.14&$-3.50\pm$0.16&$-4.92\pm$0.30&              &$0.05\pm$0.05$^a$&11.8\\
51&45.23s&38$'$54.9$''$&$-2.04\pm$0.14&$-2.78\pm$0.15&$-3.37\pm$0.15&$-4.34\pm$0.44&              &$0.16\pm$0.07$^a$&11.5\\
52&45.18s&38$'$53.0$''$&$-1.81\pm$0.15&$-2.70\pm$0.13&$-3.26\pm$0.16&$-5.06\pm$0.26&              &$0.01\pm$0.05$^a$&11.4\\
53&44.86s&38$'$57.5$''$&$-1.60\pm$0.17&$-2.59\pm$0.14&$-3.01\pm$0.23&$-4.79\pm$0.31&              &$0.01\pm$0.06$^a$&10.9\\
54&45.67s&38$'$54.7$''$&$-1.67\pm$0.15&$-2.49\pm$0.17&$-3.18\pm$0.19&$-4.77\pm$0.29&              &$0.02\pm$0.05$^a$&10.5\\
55&45.66s&38$'$53.2$''$&$-1.36\pm$0.18&$-2.27\pm$0.17&$-2.94\pm$0.19&$-4.48\pm$0.42&              &$0.00\pm$0.07$^a$& 9.6\\
56&44.96s&38$'$55.7$''$&$-1.30\pm$0.19&$-2.09\pm$0.19&$-2.84\pm$0.20&$-4.43\pm$0.48&              &$0.00\pm$0.08$^a$& 9.0\\

\end{tabular}
\caption{Positions, reddened absolute magnitudes, $E_{B-V}$ coefficients, and
initial masses of the brightest stars of \object{NGC 588}. The extinction is
discussed in \S~\ref{sect_ebv_num}, and the initial masses, in
\S~\ref{meanimf}. $^a$Values eventually set to 0.11}
\label{tab_phot}
\end{table*}

The star observed with STIS is star \#1 from Table~\ref{tab_phot}. From the
stars with available $M_{469}$, star \#2 is the only one exhibiting a clear
He{\sc II} excess, and was identified as the only or main WN star responsible
for the He{\sc II} bump in the optical spectrum.

\subsection{Identification of the star observed with STIS}
\label{stis_id}
\begin{figure}
\resizebox{\hsize}{!}{
\includegraphics[angle=90]{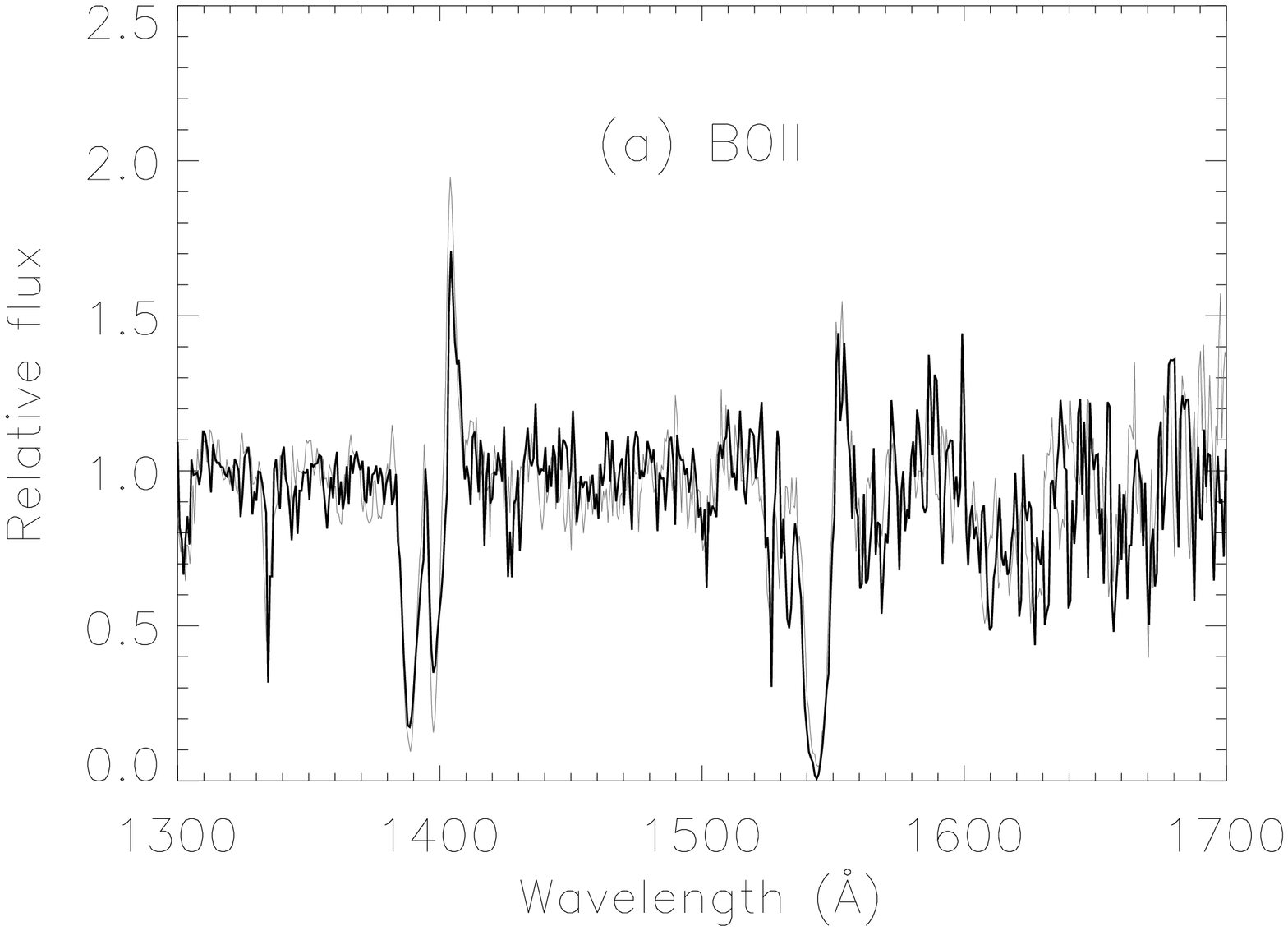}
\includegraphics[angle=90]{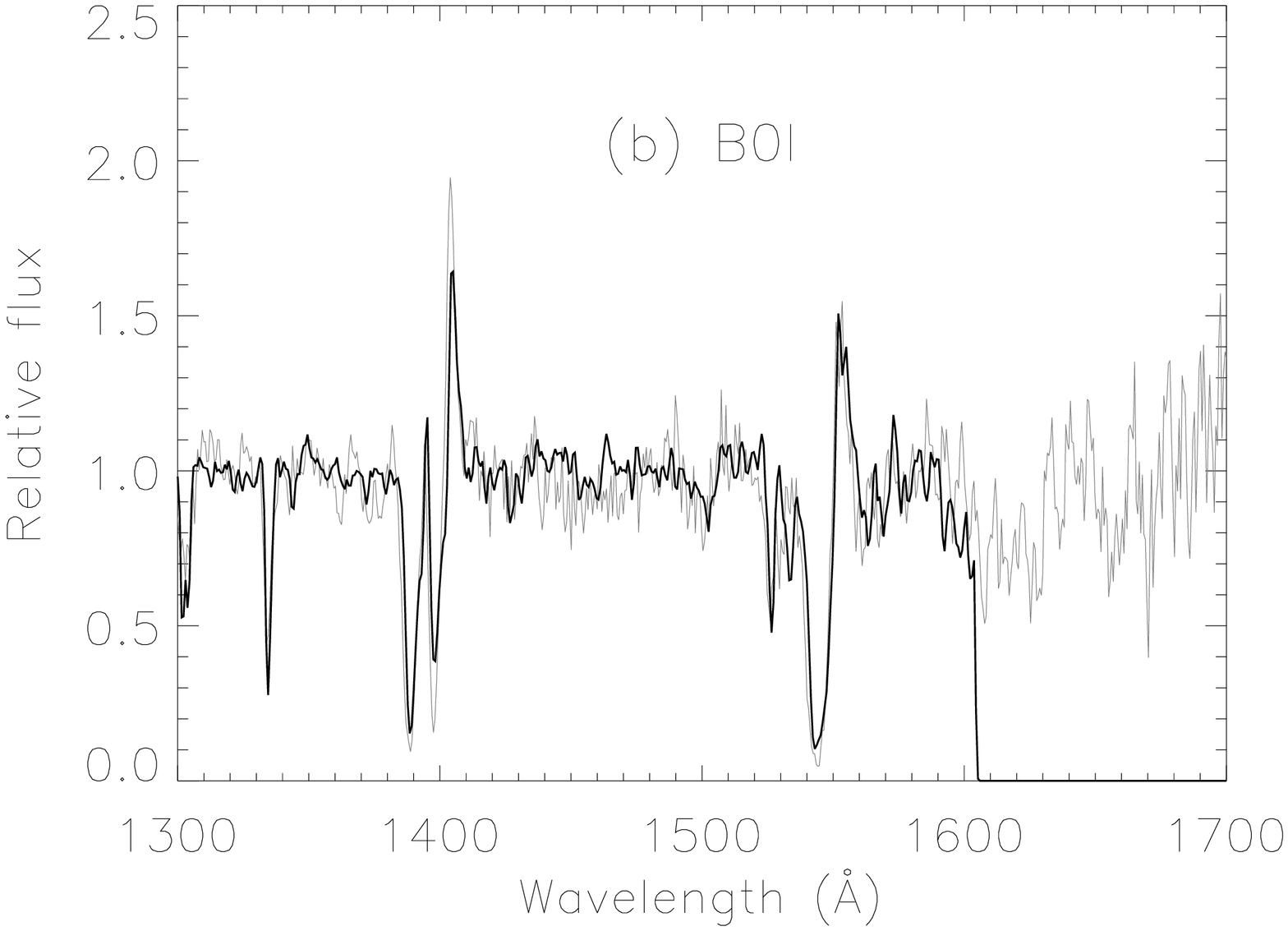}}
\resizebox{\hsize}{!}{
\includegraphics[angle=90]{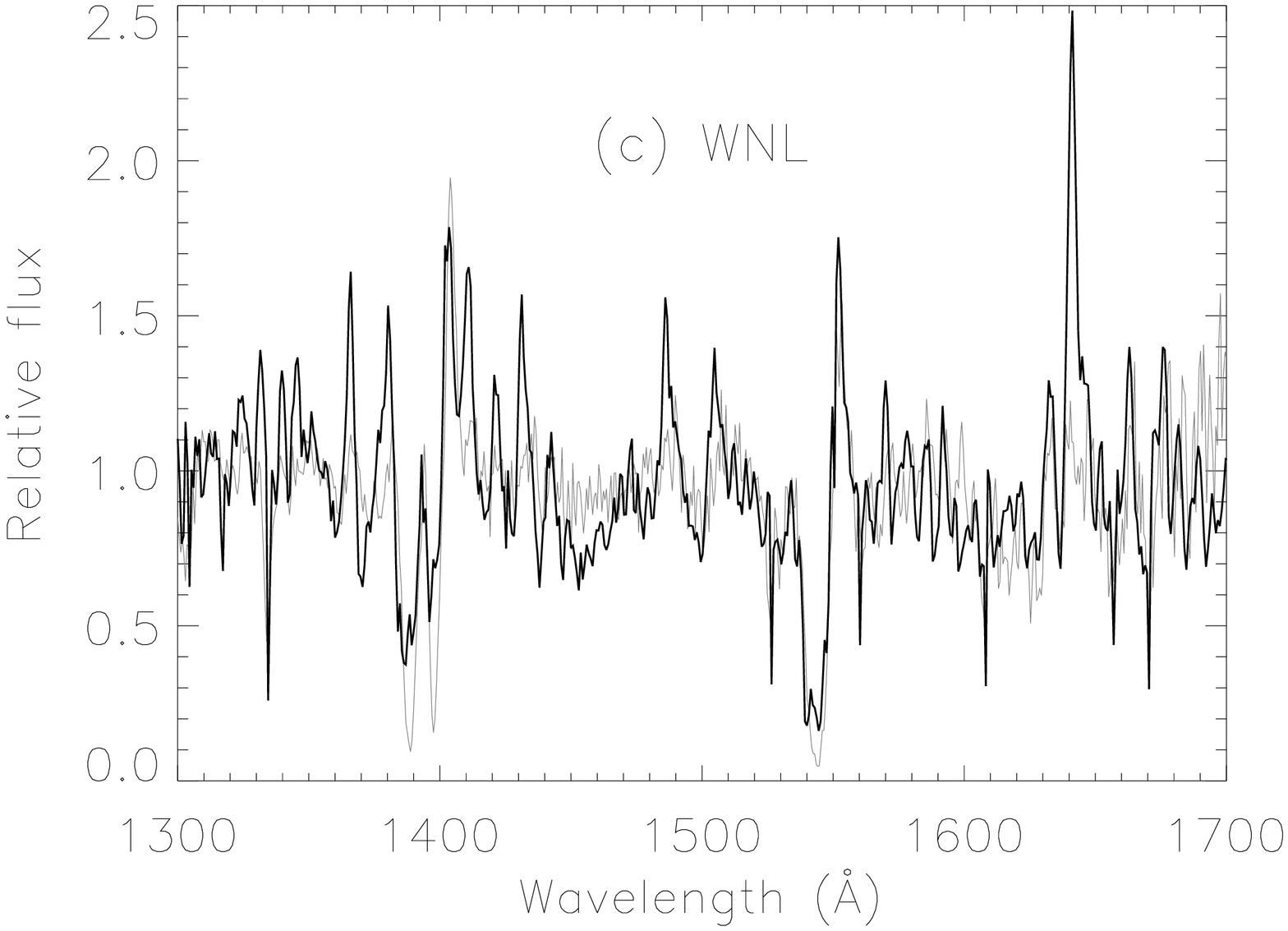}
\includegraphics[angle=90]{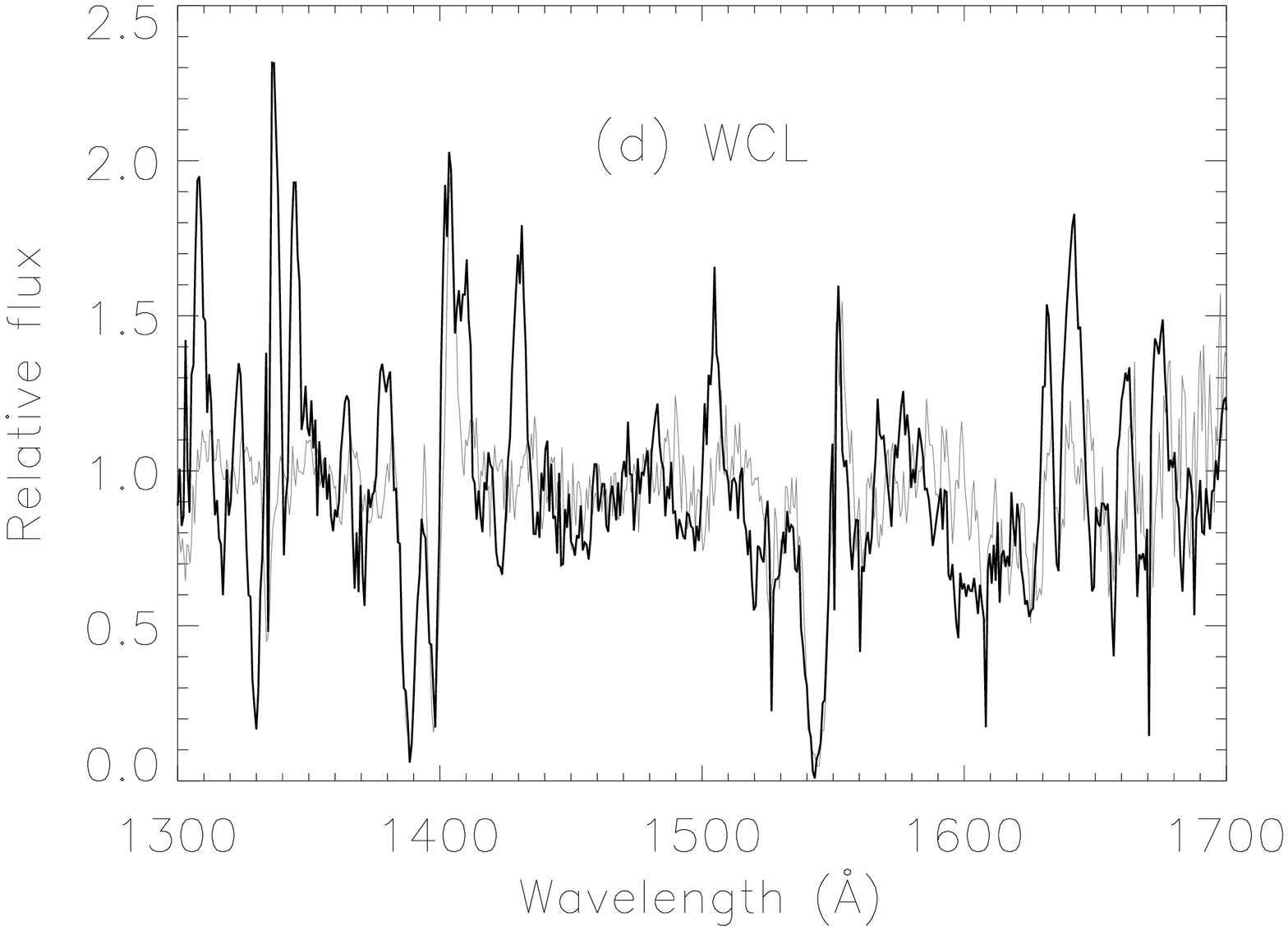}}
\caption{STIS vs. LMC/SMC spectra: B0II, B0I, WNL and WCL, respectively. The
1600--1700 {\AA} region was unavailable for the B0I star.}
\label{fig_stis}
\end{figure}
\begin{figure}
\resizebox{\hsize}{!}{\includegraphics[angle=90]{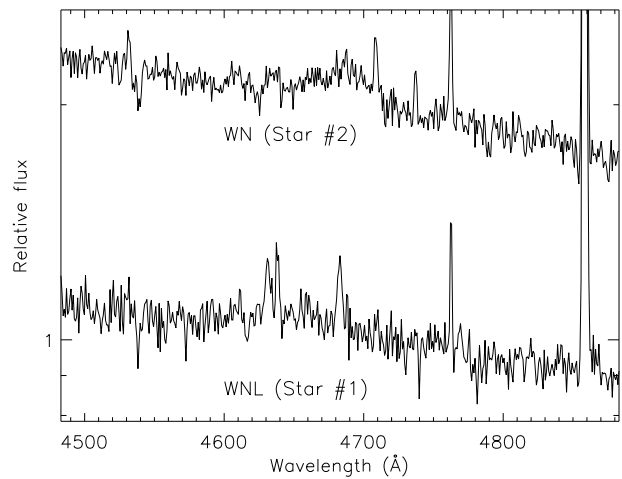}}
\caption{Optical spectra near the WR He{\sc II} $\lambda$4686 bump, extracted
from the blue CAHA spectrum around stars \#1 and \#2 of Table~\ref{tab_phot}.}
\label{fig_wrbumps}
\end{figure}

To identify the nature of the star observed with STIS, we first compared its
spectrum with the ones of the LMC/SMC library of \cite{llh01}, paying attention
to the Si{\sc IV} $\lambda$1400, C{\sc IV} $\lambda$1550 and, when available,
He{\sc II} $\lambda$1640 lines. The four acceptable matches
(Fig.~\ref{fig_stis}) indicated the star to be of class B0II, B0I, late WN
(WNL) or late WC (WCL). We found a significant He{\sc II} $\lambda$4686 bump in
the CAHA slit zone situated within the PSF extent of this quite isolated star
(see Fig.~\ref{fig_wrbumps} for this star and for star \#2 of
Table~\ref{tab_phot}). The intensity of this bump being compatible with the
(quite inaccurate) value derived from the photometric measurements, we deduced
that the star is a WR. In absence of any optical carbon feature, in contrast
with the well-defined He{\sc II} $\lambda$4686 bump, we concluded that it is a
WNL star.

Considering the presence of such a star in the cluster, we inferred the age
range of the latter for both metallicities Z=0.004 and Z=0.008: 3.2--3.7 and
2.8--4.5 Myr, respectively.

\subsection{Comparison with isochrones}
\subsubsection{Construction of the isochrones}
We constructed isochrone curves calling the same evolutionary tracks and model
atmospheres as in Section~\ref{classi}. The effective temperatures and
bolometric luminosities of these isochrones were computed with the original
Starburst99 code. The corresponding magnitudes, however, were derived from them
by means of an interpolation method different from the one of Starburst99. The
latter mainly consists in nearest-neighbor selection of the spectra in the
$(\log T_{\rm eff},\log g)$ plane, $T_{\rm eff}$ being the effective
temperature, and $g$, the surface gravity. This induces step-like
discontinuities in plots such as color-magnitude diagrams. This is why we
exploited an alternative interpolation method, explicated in the Appendix, that
consists, at each wavelength, of applying a bilinear interpolation of the
logarithm of the flux in the $(\log g,y)$ plane, $y$ being a variable that
depends both on the effective temperature and on the wavelength.

Depending on the stellar parameters, various kinds of model atmospheres, listed
in Section~\ref{classi}, were exploited. This implies an artificial jump in the
isochrone diagrams at the locus of transition from \cite{lbc97} models to the
one of \cite{snc02}. Fortunately, this transition occurs in a zone where it is
smaller than the error bars, and consequently it has a negligible impact.

\subsubsection{Reddening}
\label{sect_ebv_num}
\begin{figure}
\resizebox{\hsize}{!}{
\includegraphics{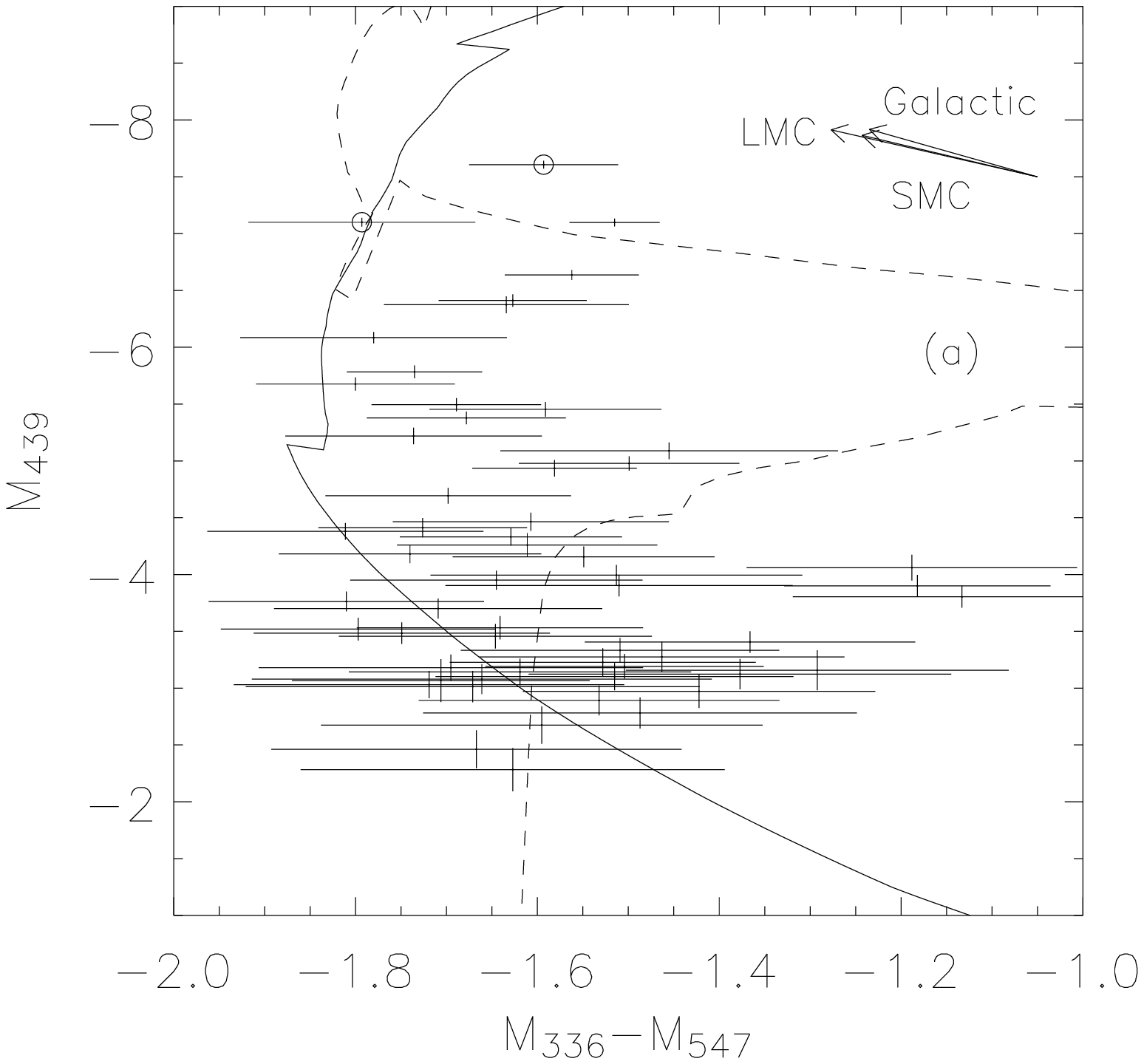}
\includegraphics{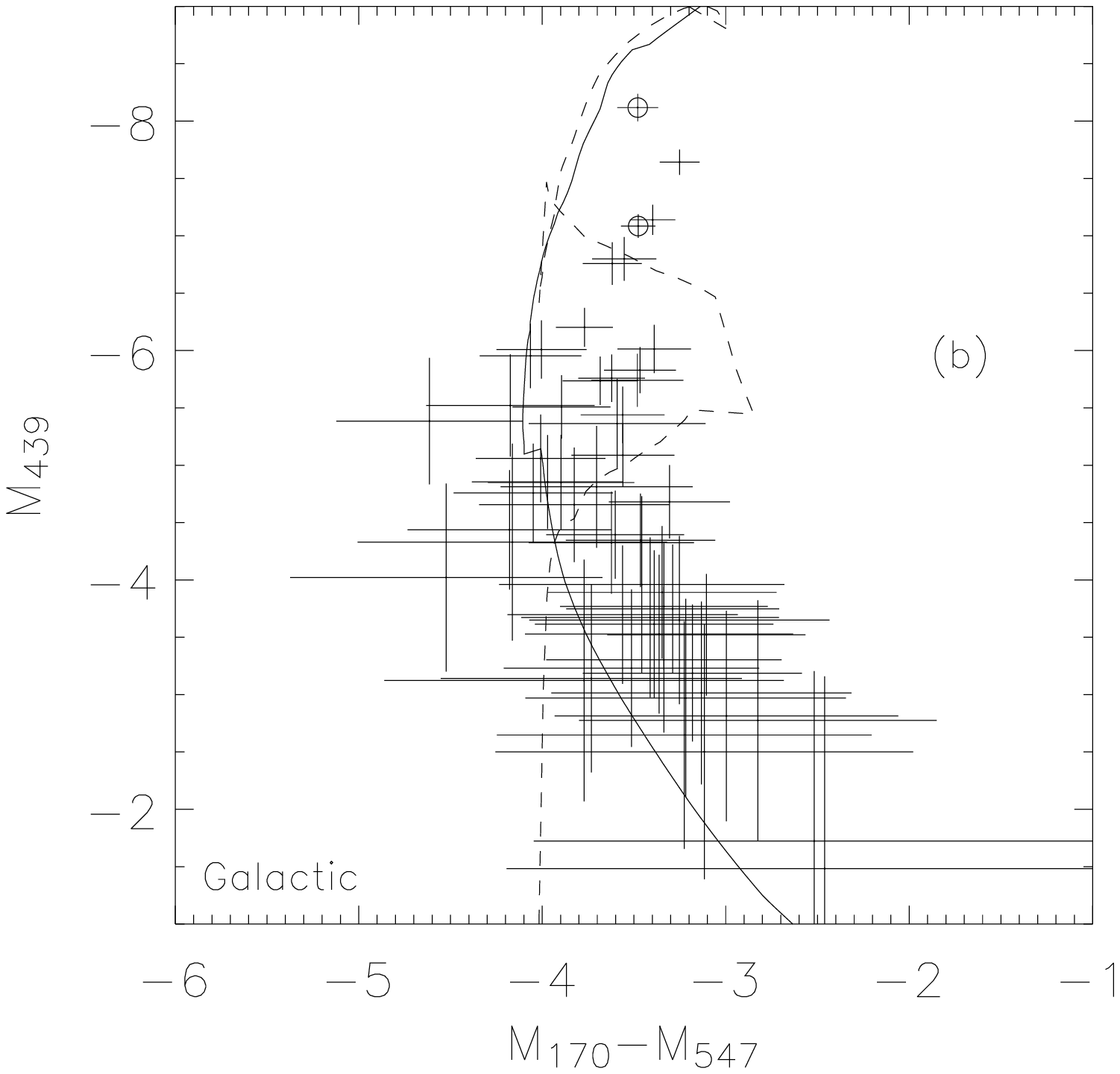}}
\resizebox{\hsize}{!}{
\includegraphics{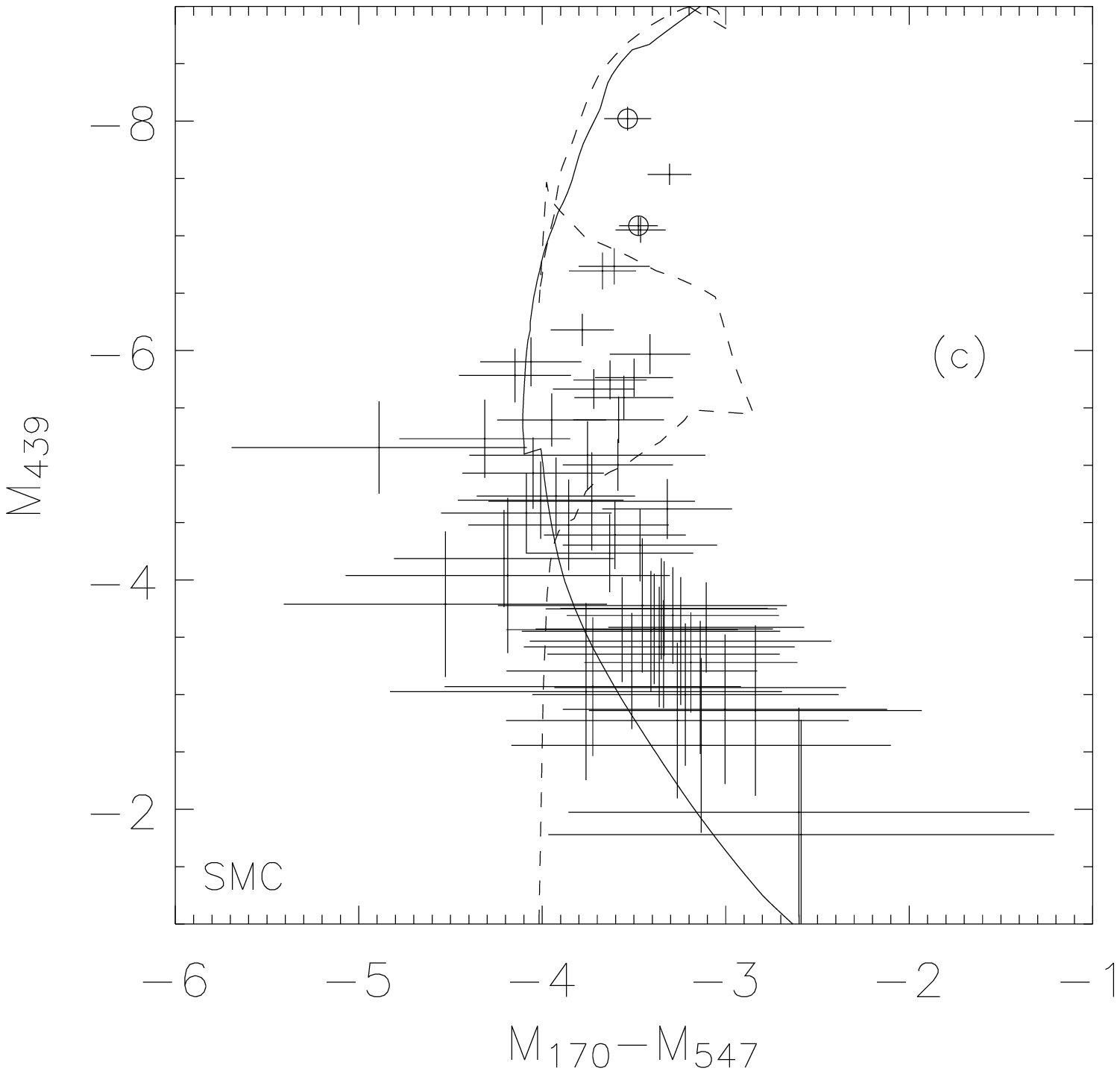}
\includegraphics{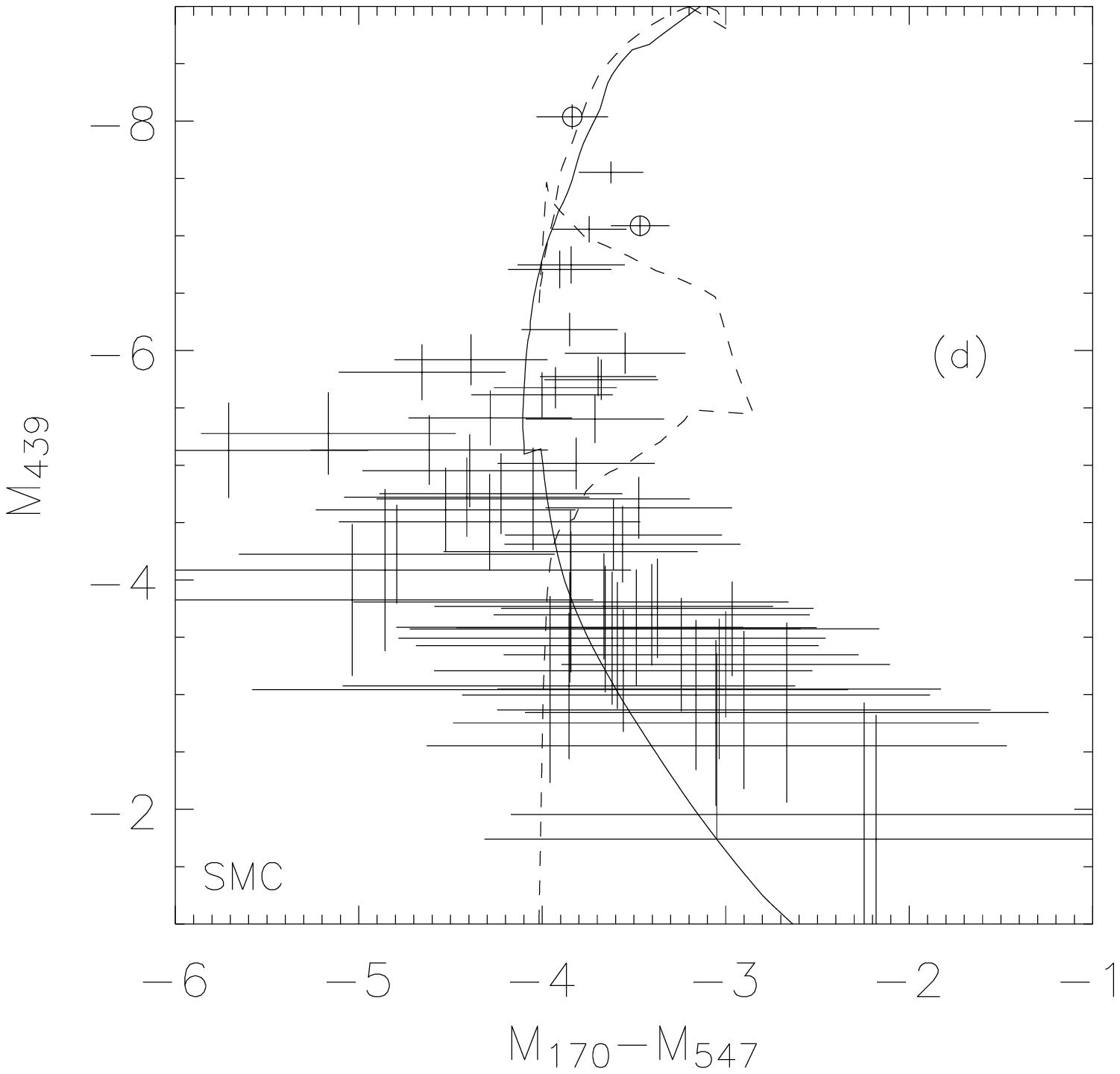}}
\caption{Illustration of the dereddening procedure. The isochrone shown here
was established for an age of 3.5 Myr and for Z=0.008. Panel (a): observed
$(M_{439},M_{336}-M_{547})$ diagram. Other panels: $(M_{439},M_{170}-M_{547})$
diagram dereddened with the $(M_{439},M_{336}-M_{547})$ one and the Galactic
(b), LMC (c) and SMC (d) laws. The dotted lines are the WR branches of the
isochrone curves. The two WR stars are signalized by open circles. The arrows
indicate the shift that the observational points undergo by if dereddened for
$E_{B-V}=0.1$. In the diagrams of panels (b), (c) and (d), the error bars
account for uncertainties of $E_{B-V}$.}
\label{fig_dered_one}
\end{figure}
\begin{figure}
\resizebox{\hsize}{!}{
\includegraphics{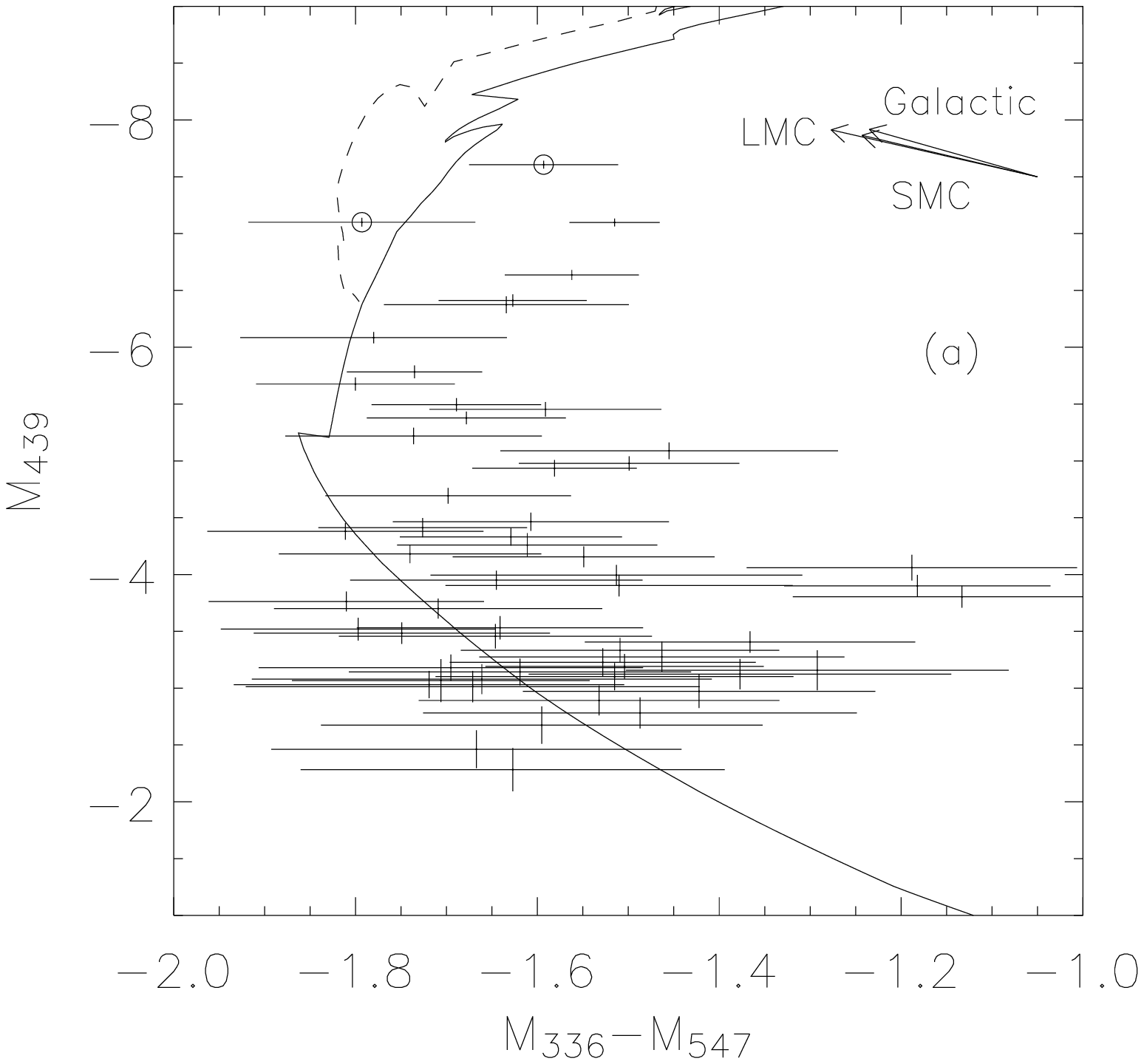}
\includegraphics{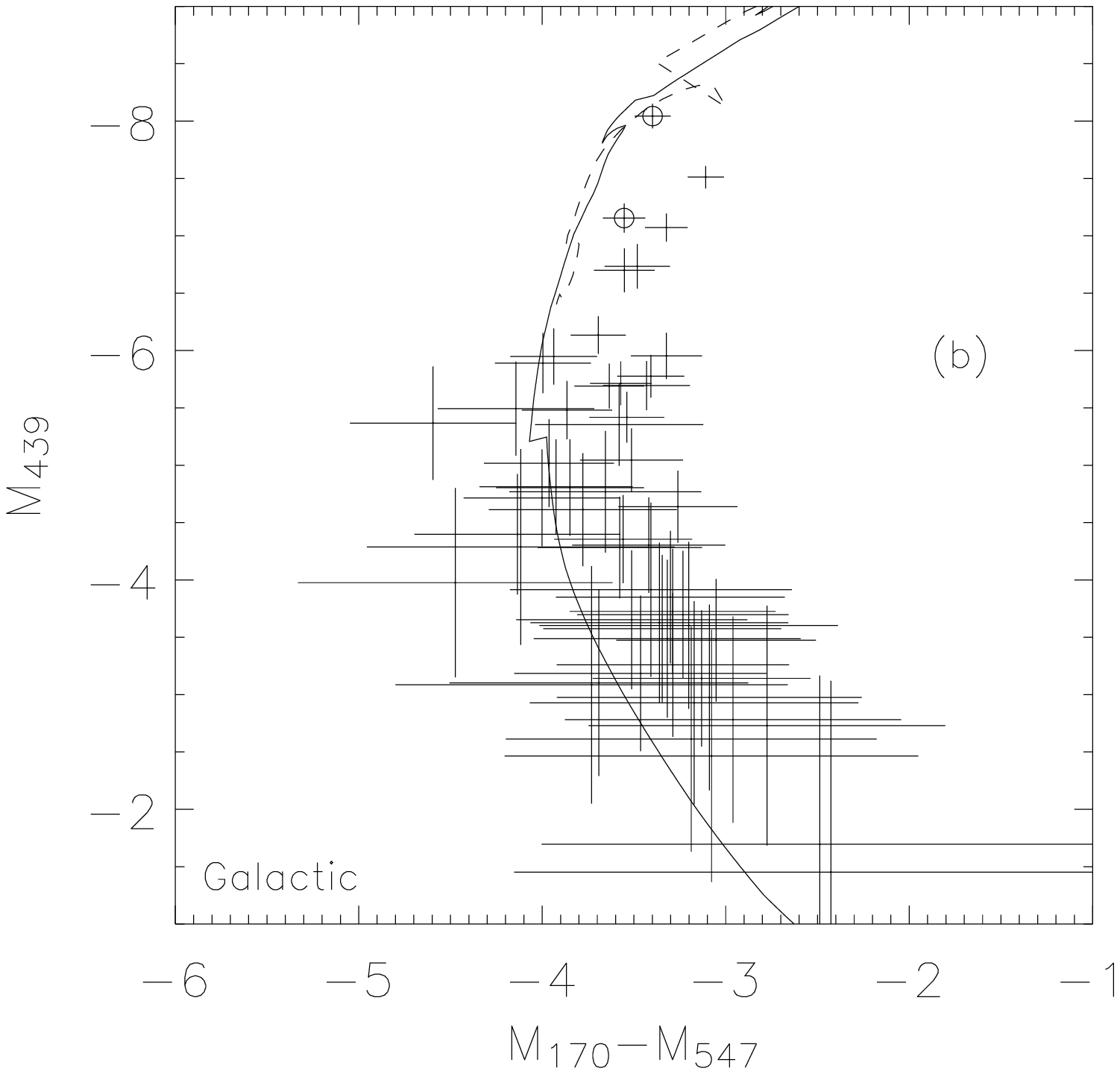}}
\resizebox{\hsize}{!}{
\includegraphics{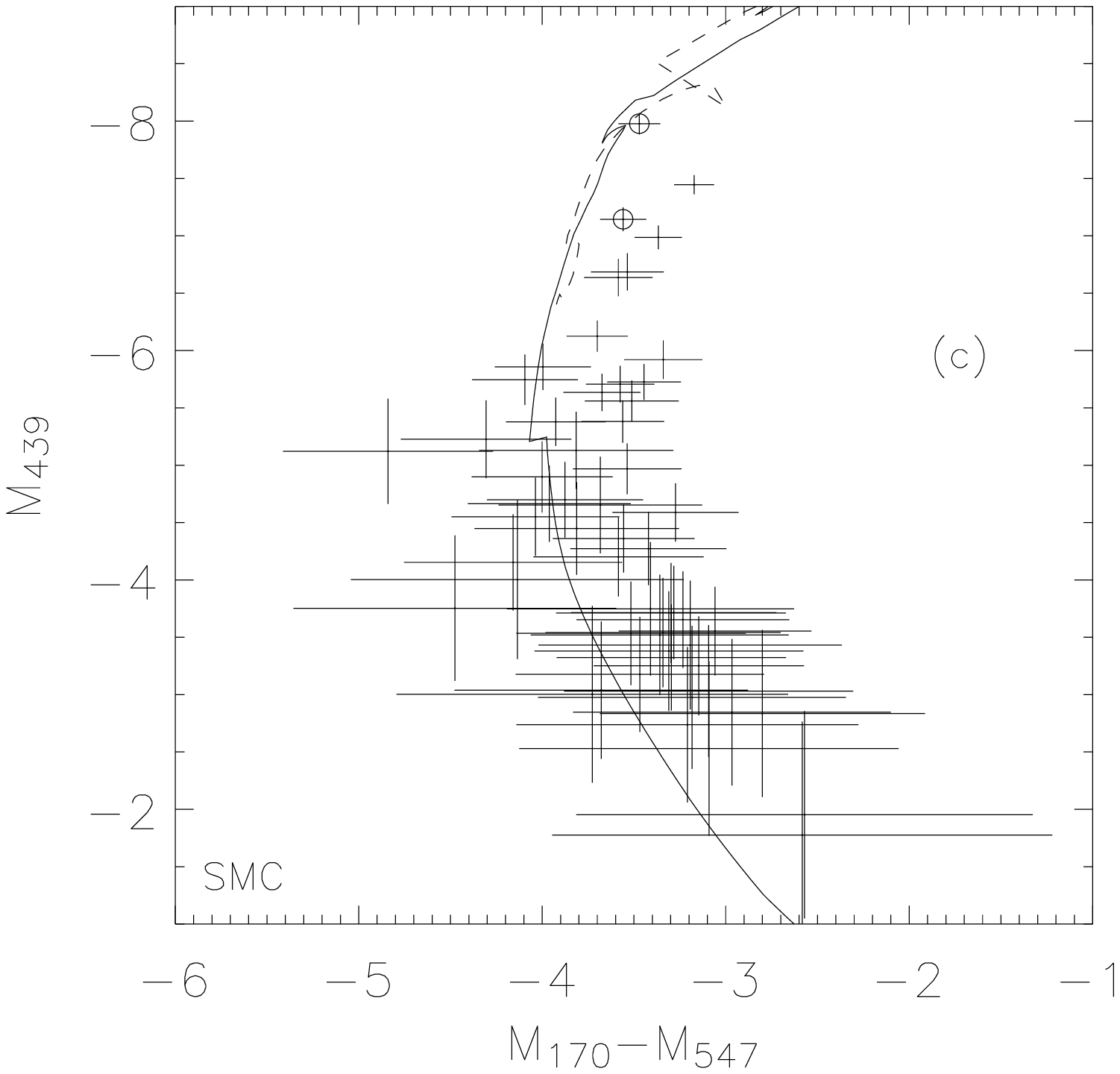}
\includegraphics{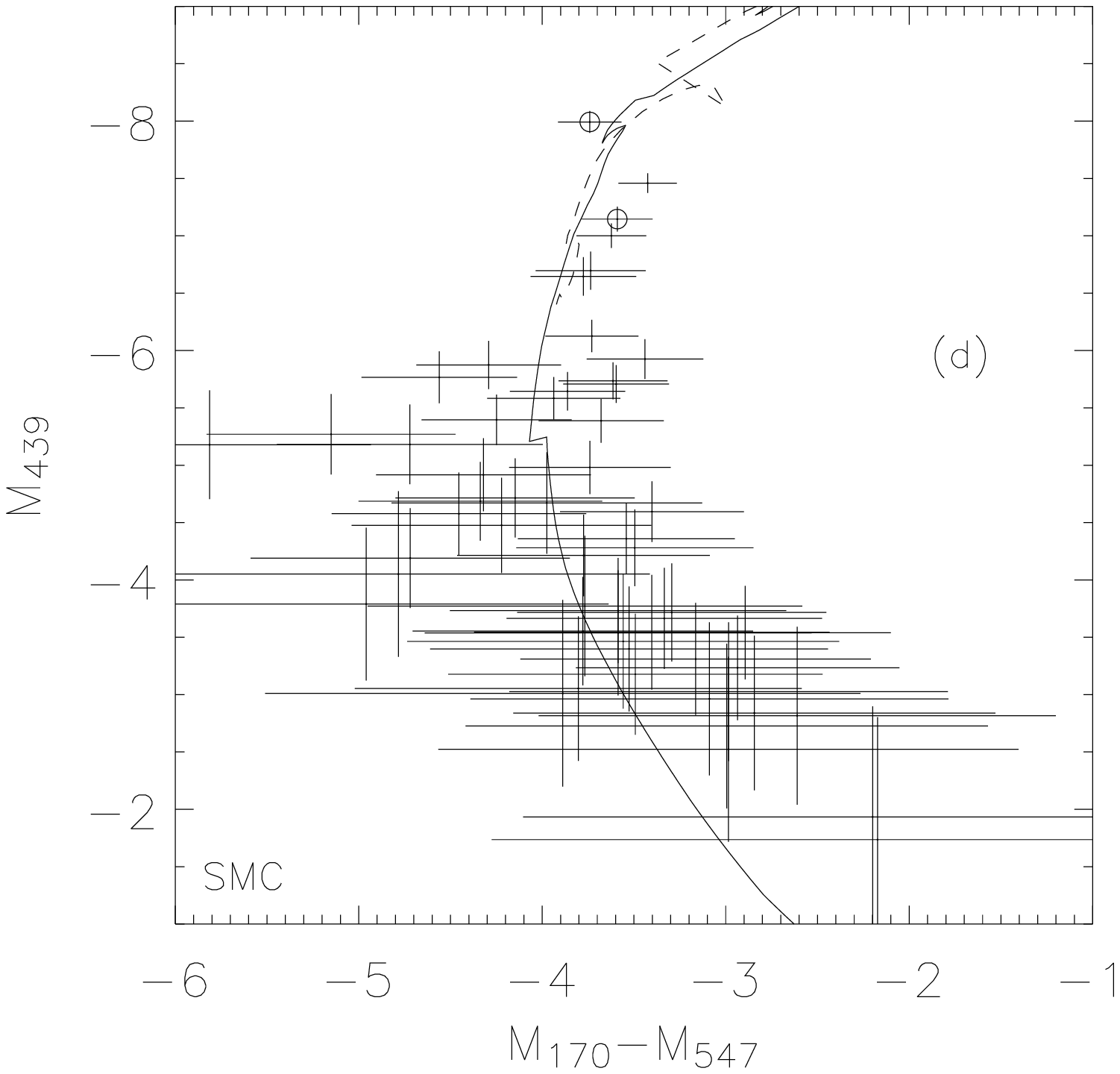}}
\caption{Same as Fig.~\ref{fig_dered_one} for the isochrone of 4.5 Myr.}
\label{fig_dered_two}
\end{figure}
\begin{figure}
\resizebox{\hsize}{!}{\includegraphics[angle=90]{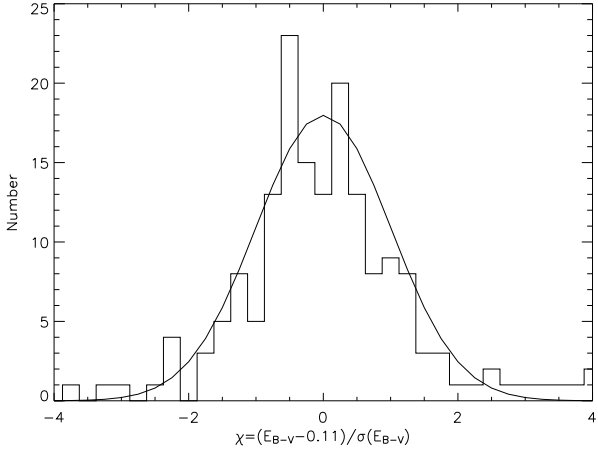}}
\caption{Histogram of the dispersion of the stellar color excesses $E_{B-V}$
around the nebular one 0.11. The abscissa axis variable is the discrepancy term
$\chi=(E_{B-V}-0.11)/\sigma(E_{B-V})$, where $\sigma(E_{B-V})$ is the
uncertainty of the considered measurement of $E_{B-V}$. The histogram includes
the 173 stars observed simultaneously through filters F547W and F439W. The
overplotted curve is the theoretical expected distribution in the case of a
unique real value $E_{B-V}=0.11$ and of well-determined measurement
uncertainties.}
\label{hist_ebv}
\end{figure}

We first investigated the extinction law, since a priori it may be different
from the one inferred from the canonical study of the cluster (cf.
\S~\ref{sedcol}). We considered several isochrones in the age range of the
cluster, and for each of them and each of the three tested extinction laws
(Galactic, LMC and SMC), we proceeded as follows. We first estimated the color
excess of each star in Table~\ref{tab_phot} by shifting the measured points
towards the isochrone curve in the $(M_{439},M_{336}-M_{547})$ diagram
according to the tested law, assuming stars \#1 and \#2 to be WN stars, and all
the other ones to be on the main sequence. Then, we dereddened each star in the
$(M_{439},M_{170}-M_{547})$ diagram, and compared the resulting points with the
theoretical isochrone. Independently of the reference isochrone (i.e., in
practice, the age), the SMC law was favored, as the systematic discrepancy
between the isochrone and the locus of the stars, in particular the brightest
ones, was small for the SMC law and significantly larger for the other two
laws. Numerically, the discrepancy between the observational points in the
$(M_{439},M_{170}-M_{547})$ diagram dereddened with the use of the
$(M_{439},M_{336}-M_{547})$ one and the theoretical isochrone is characterized
by a reduced $\chi^2$ ranging in the approximate intervals 2.3--3.3 for the
Galactic law, 1.8--2.0 for the LMC one and 0.6--0.8 for the SMC one, obtained
with 56 stars. The fact that this $\chi^2$ is close to 1 with the SMC law made
us confident in the latter, and is also an indication that none of the 56
considered stars is significantly affected by blending with an unresolved
companion. The procedure of selection of an extinction law is illustrated in
Fig.~\ref{fig_dered_one} for an age of 3.5 Myr, and in Fig. \ref{fig_dered_two}
for 4.5 Myr.

Once the extinction law chosen, we re-considered the individual color excess
$E_{B-V}$ of the stars. For the stars used in the selection of the extinction
law, we started from the values already estimated in the
$(M_{439},M_{336}-M_{547})$ diagram. For the stars that, once dereddened, were
found to be brighter than $M_{439}=-6$, we conserved the values of $E_{B-V}$
found in this diagram and the corresponding uncertainties, the latter including
the (small) scatters of values found with the different possible isochrones.
However, for the stars dimmer than $M_{439}=-6$, we used the constancy of the
isochrones in the $(M_{439},M_{170}-M_{547})$ diagram to estimate $E_{B-V}$
with better accuracy. Indeed, the uncertainty in $E_{B-V}$ determined in one of
the $(M_{439},M_X-M_{547})$ diagrams ($X=170$ or $X=336$) is roughly the ratio
$\sigma(M_X-M_{547})/(f_X-f_{547})$, where $\sigma(M_X-M_{547})$ is the
uncertainty in the $M_X-M_{547}$ color, and in our case, this ratio in
$E_{B-V}$ is smaller for $X=170$ than for $X=336$. With a few exceptions, we
found all the color excess to be compatible with the nebular value,
$E_{B-V}=0.11$, and when this was the case and the dereddened F439W magnitude
was dimmer than $-6$, we set them to this value. An important departure from
$E_{B-V}=0.11$ is the case of star \#2: $E_{B-V}=0.00\pm 0.03$. All stars with
unavailable F170W and F336W magnitudes were dereddened with $E_{B-V}=0.11$,
compatible with the $(M_{439},M_{439}-M_{547})$ theoretical curve.
Fig.~\ref{hist_ebv} testifies to the correctness of setting almost all the 173
stellar color excesses to the nebular one.

In all what follows, the color-magnitude diagrams (CMDs) refer to the
dereddened stellar magnitudes.

\subsubsection{Fit of the age and of the metallicity}
\begin{figure}
\resizebox{\hsize}{!}{
\includegraphics{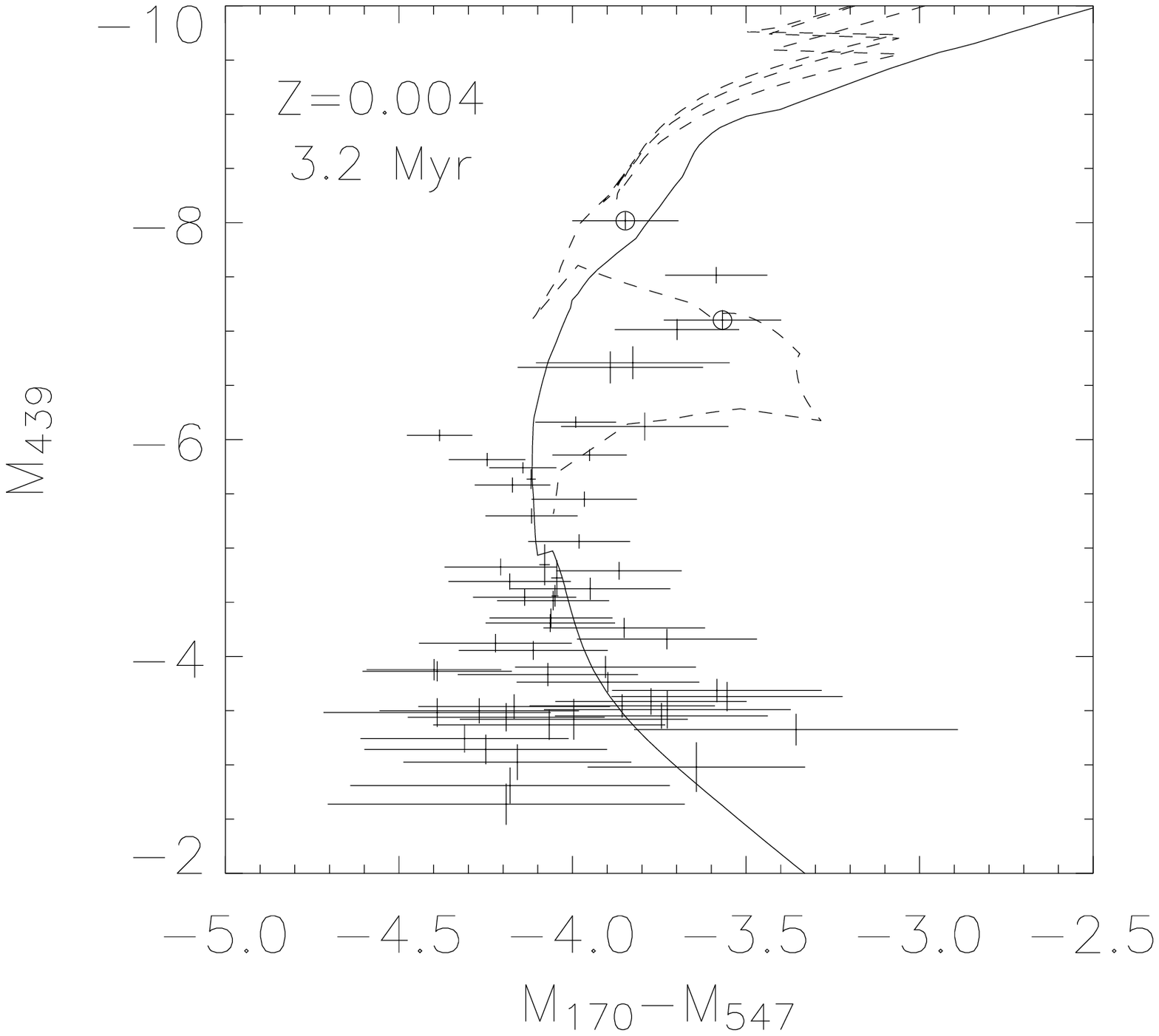}
\includegraphics{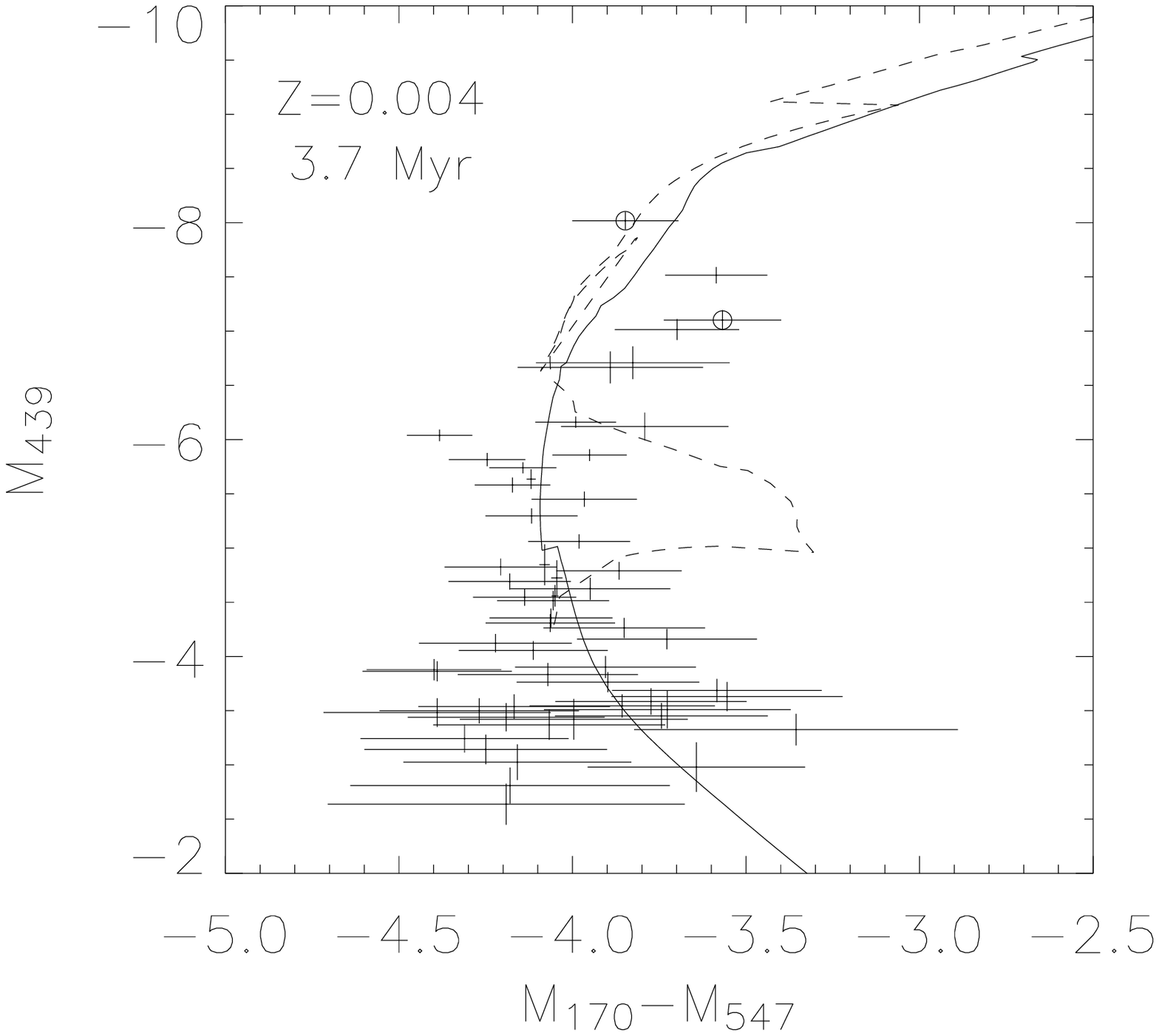}}
\caption{Observed vs. isochrone $(M_{439},M_{170}-M_{547})$ diagram for
Z=0.004. The two WR stars are marked with open circles. The WR branches of the
isochrones are shown as dashed lines.}
\label{fig_isoc_004}
\end{figure}
\begin{figure}
\resizebox{\hsize}{!}{
\includegraphics{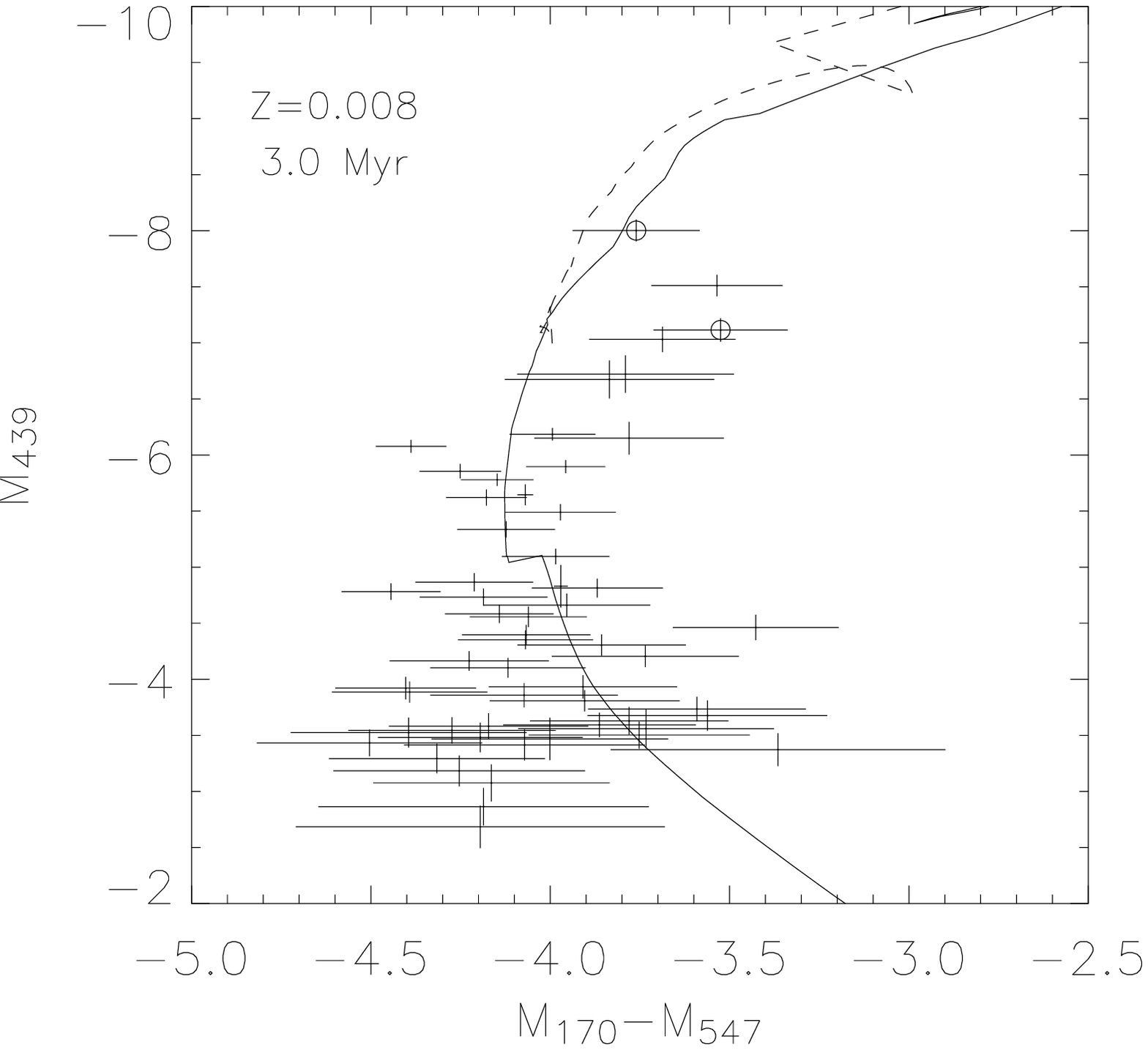}
\includegraphics{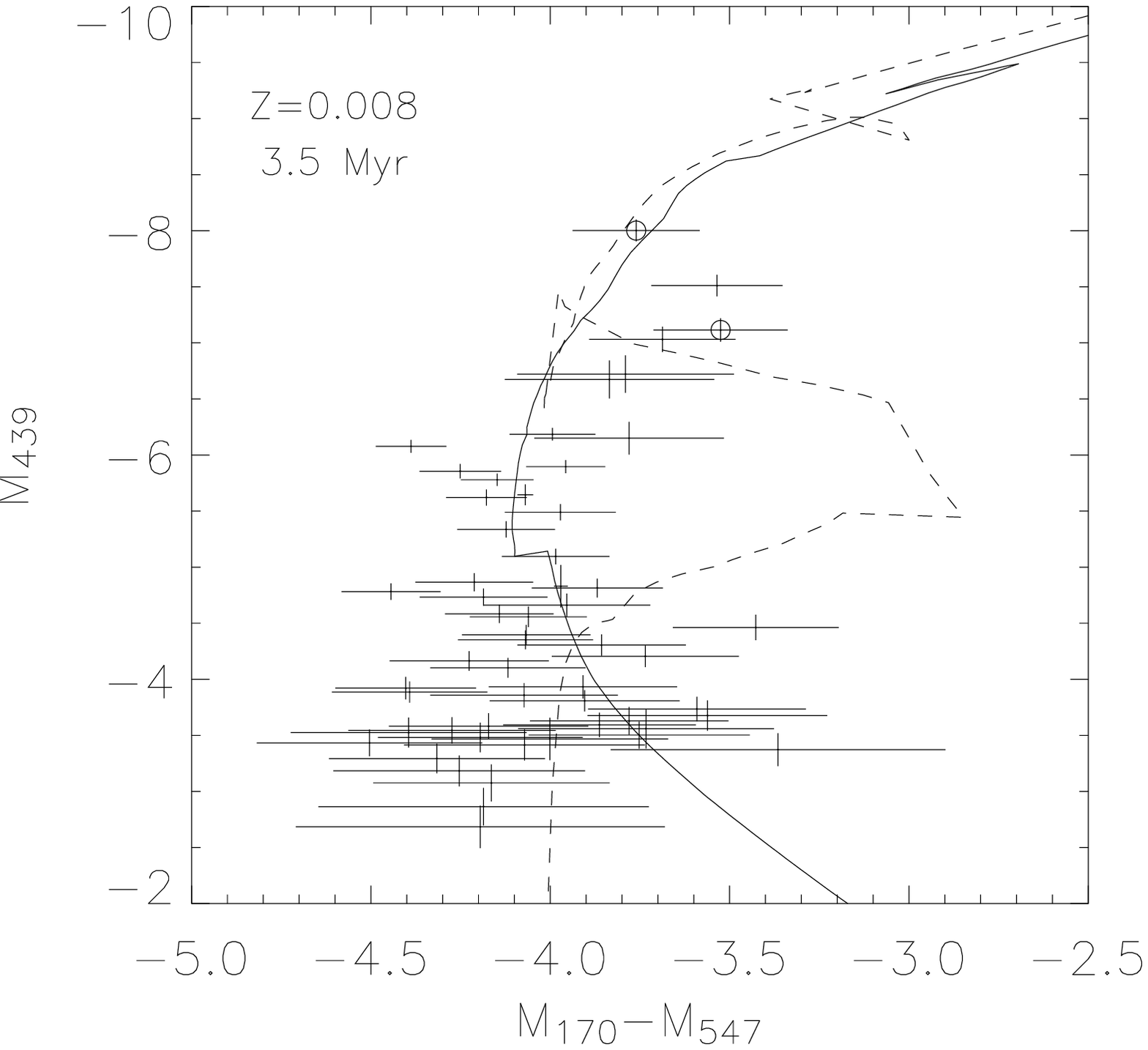}}
\resizebox{\hsize}{!}{
\includegraphics{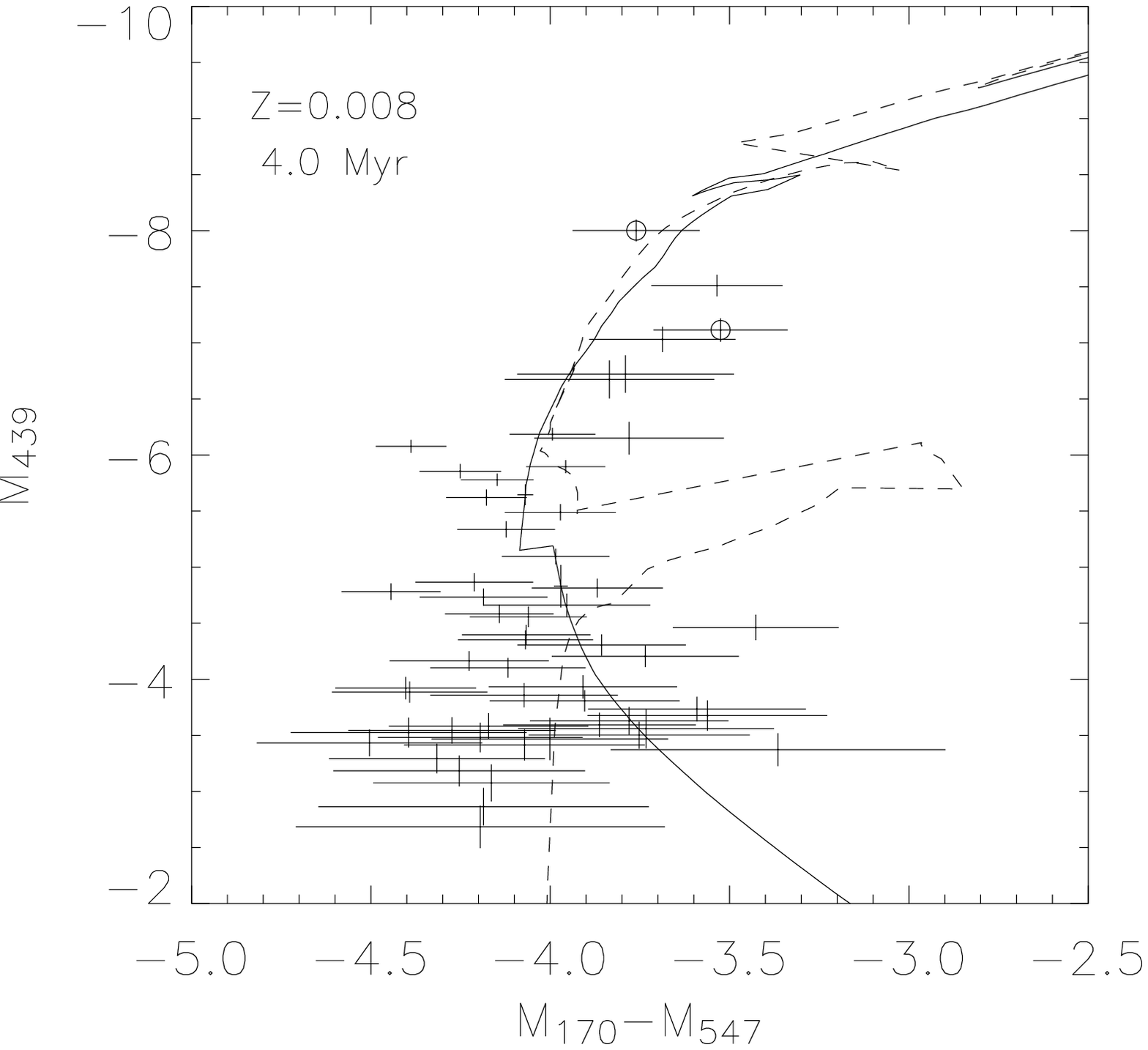}
\includegraphics{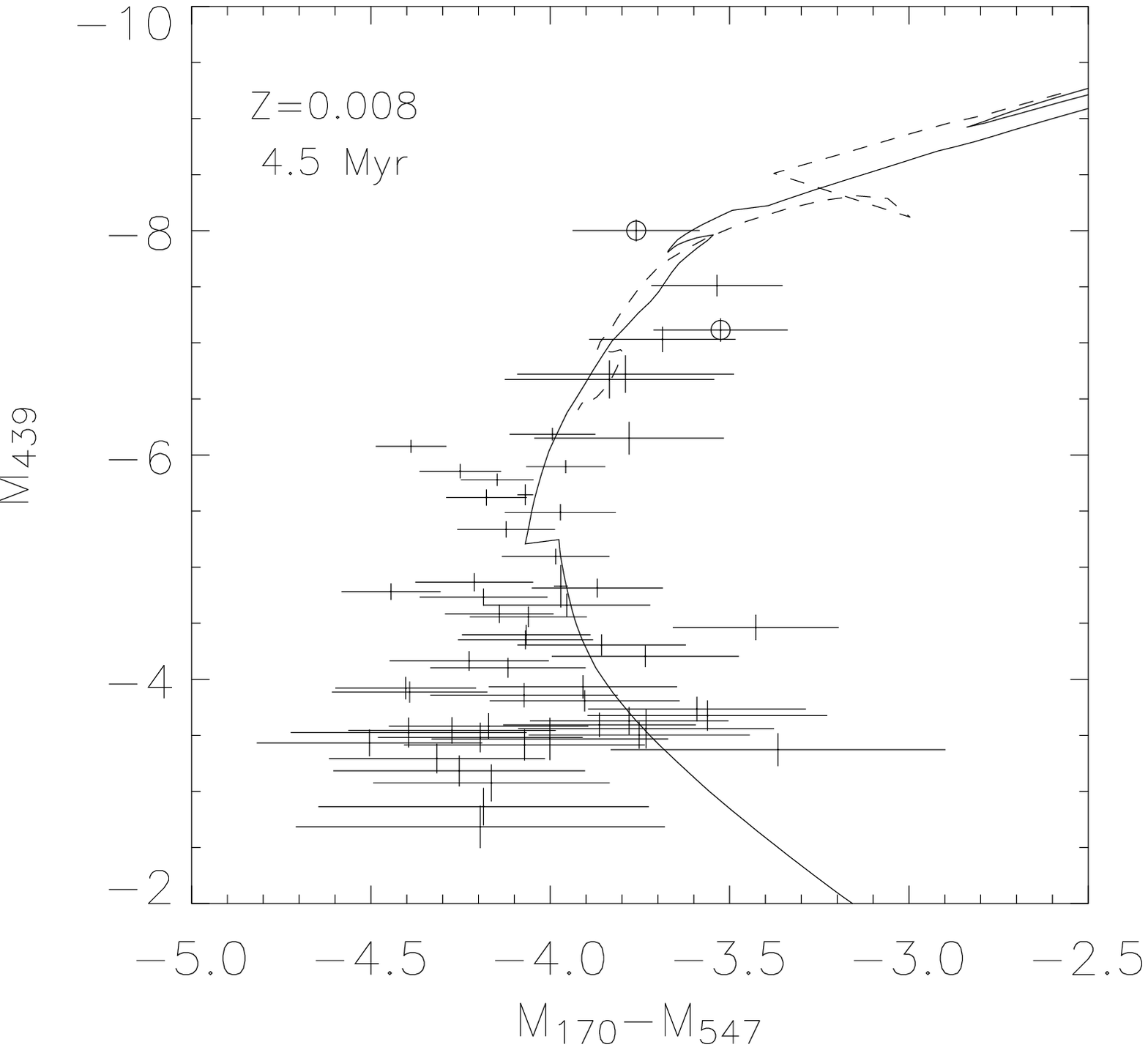}}
\caption{Same as Fig.~\ref{fig_isoc_004} for Z=0.008.}
\label{fig_isoc_008}
\end{figure}
\begin{figure}
\resizebox{\hsize}{!}{
\includegraphics{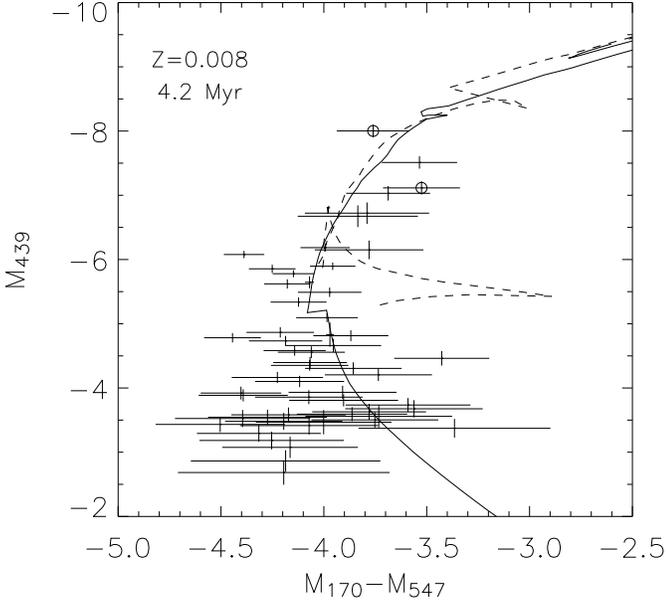}}
\caption{Final adopted $(M_{439},M_{170}-M_{547})$ diagram.}
\label{finaldiag}
\end{figure}
\begin{figure}
\resizebox{\hsize}{!}{
\includegraphics{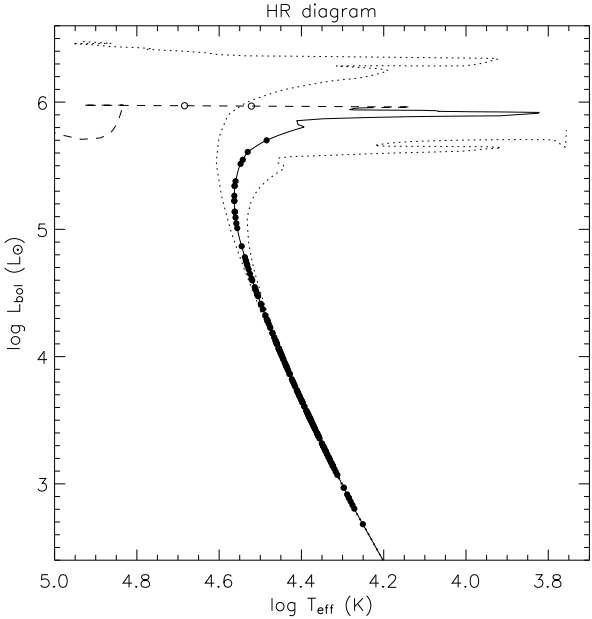}}
\caption{Final adopted HR diagram for all the detected stars. The circles show
the selected individual stellar model. The dashed part of the diagram is the WR
branch. The two dotted curves represent the HR diagram at 3.0 Myr (upper curve)
and 5.5 Myr (lower curve)}
\label{hrdiag}
\end{figure}
The age and metallicity of the cluster were constrained by simultaneously
fitting the isochrone to the dereddened observational points in the
$(M_{439},M_{170}-M_{547})$ diagram, and by comparing synthetic spectra and
derived properties to the observations. For a given isochrone, each star was
identified to the nearest point of the isochrone in terms of chi-square (i.e.,
accounting for the error bars) in the $(M_{439},M_{170}-M_{547})$ diagram or,
when the F170W magnitude was unavailable, in the $(M_{439},M_{439}-M_{547})$
diagram, thus attributing the most appropriate model spectrum to the considered
star. If the cluster spectrum was synthesized to simulate what is seen through
a slit (CAHA or IUE), the spectrum selected for a given star was multiplied by
its corresponding aperture throughput (known from its position with respect to
the slit and from the angular PSF or seeing of the considered data) and a
function representing its possible differential extinction (with respect to the
reference value $E_{B-V}=0.11$) along the processed wavelength range, and added
to the total cluster spectrum to compute. If the total cluster spectrum was to
be synthesized, then the individual stellar spectra were summed without being
previously modified.

We computed the isochrone curves for the different ages multiple of 0.1 Myr
belonging to the ranges constrained by the presence of a WNL star. Here, we
show the curves obtained for the ages 3.2 and 3.7 Myr in the case Z=0.004
(Fig.~\ref{fig_isoc_004}), and 3.0, 3.5, 4.0 and 4.5 Myr for Z=0.008
(Fig.~\ref{fig_isoc_008}). At both metallicities, the isochrone high-luminosity
main-sequence branch was found to move, with increasing age, from regions of
low values of the $M_{170}-M_{547}$ color towards the somewhat ``red''
observational points.

The synthesized spectra were exploited as follows. For a given tested
isochrone, the spectrum {\bf simulated for the optical slit} was used to
extract the model ratio $R_B=F_\lambda(3630)/F_\lambda(3780)$ related to the
Balmer jump, which is a diagnostic of the effective temperature of a cluster
dominated by hot stars. Indeed, in the spectrum of a hot star, the intensity of
the Balmer jump decreases with increasing effective temperature, and usually
serves to determine the subtypes of B stars, whereas it is negligible in the
spectra of O stars. The observed value, whose uncertainty is dominated by the
local fluctuating residues in photometric calibration, is $R_B=1.12\pm0.08$.
The constructed UV spectrum was visually compared to the IUE observed one.
Finally, we derived the predicted $W({\rm H}\beta)$ from the total cluster
spectrum, assuming again that the nebula absorbs all the ionizing photons, and
thus requesting the predicted $W({\rm H}\beta)$ to be greater than or equal to
the observed one.

Mathematically, we used the following $\chi^2$ estimator to constrain the age
and the metallicity of the cluster:
\begin{eqnarray}
\chi^2&=&\left(\frac{\max(330-W({\rm H}\beta),0)}{30}\right)^2+\left(\frac{R_B-
1.12}{0.08}\right)^2\nonumber\\+\chi_{\rm isoc}^{2}
\end{eqnarray}
where $\chi_{\rm isoc}^{2}$ characterizes, for a given isochrone, the
discrepancy between the latter and the loci of the brightest four MS stars in
the $(M_{439},M_{170}-M_{547})$ diagram. The numerical results are summarized
in Table~\ref{tab_res_num}. In practice, the synthetic Balmer jump was
constant, and the fit was constrained by $W({\rm H}\beta)$ and $\chi_{\rm
isoc}^{2}$. For Z=0.004, even at 3.7 Myr, the theoretical isochrone remained
toward values of $M_{170}-M_{547}$ lower (``bluer'') than the observational
points of the brightest stars, though the fit was acceptable. Meanwhile, for
Z=0.008, the fit of the isochrone was very satisfactory for ages around 4.0
Myr, with $W({\rm H}\beta)$ also being well predicted. We retained the model
with Z=0.008 and $\tau=4.2$ Myr as the best-fit one. The
$(M_{439},M_{170}-M_{547})$ diagram for this model is shown in
Fig.~\ref{finaldiag}, and the corresponding Hertzsprung-Russel (HR) diagram, in
Fig.~\ref{hrdiag}. According to the classification of \cite{sk82}, 20 stars
were classified as O-type ones.

\begin{figure}
\resizebox{\hsize}{!}{\includegraphics[angle=90]{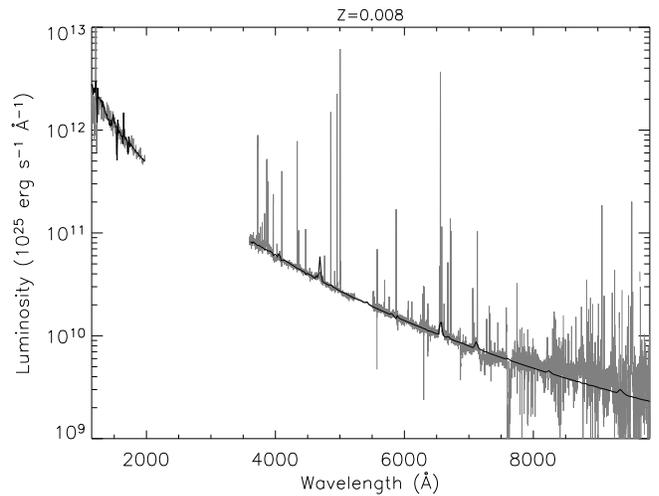}}
\caption{Optical and UV SEDs synthesized for Z=0.008 and $\tau=4.2$ Myr with
the individual stellar models. For better visibility and comparison with
Fig.~\ref{an_sed}, the luminosities of the observed optical and UV parts were
rescaled with multiplicative factors, in order to fit the integrated
luminosities in band F439W and F170W, respectively. The synthetic spectra were
normalized using the same factors.}
\label{fig_sed_phot}
\end{figure}

\begin{figure}
\resizebox{\hsize}{!}{\includegraphics[angle=90]{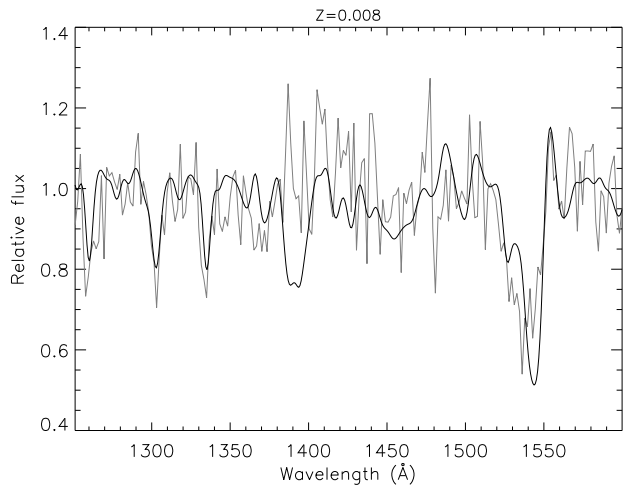}}
\caption{UV spectrum synthesized for Z=0.008 and $\tau=4.2$ Myr with the
individual stellar models.}
\label{fig_uvl_phot}
\end{figure}

The SED of the best-fit model is shown in Fig.~\ref{fig_sed_phot}. We
considered it to satisfactorily reproduce the observations, except for the
strength of the optical WR bumps. However, the model atmospheres used for WR
stars here are not intended to compute reliably these bumps, and we did not pay
attention to their strength. Instead, we estimated them from Table~1 of
\cite{sv98}, knowing that star \#2 was finally classified as a WNL from its
hydrogen surface abundance. The results, satisfying but somewhat inaccurate
(due in particular to the scatter of the model bump intensities), are
summarized in Table~\ref{tab_wr}. The synthetic UV spectrum, superimposed to
the observed one in Fig.~\ref{fig_uvl_phot}, was found to satisfactorily
reproduce the C{\sc IV} $\lambda$1550 line. It also contains an undetected
Si{\sc IV} $\lambda$1400 feature. However, the UV line spectrum is dominated by
a small number of massive stars, that are not necessarily well represented by
the spectra LMC/SMC library, in particular the WNL ones. In consequence, we did
not consider the Si{\sc IV} $\lambda$1400 discrepancy as a critical one.
\begin{table}
\begin{tabular}{cccc}
 & Z=0.004 & Observed & Z=0.008 \\
\hline
$\chi_{\rm isoc}^{2}$ & 5.2 & & 2.3 \\
$W({\rm H}\beta)$ (\AA) & 602 & 330$\pm$30 & 342 \\
$R_{\rm B}$ & 1.10 & 1.12$\pm$0.08 & 1.10 \\
\hline
age (Myr) & 3.7 (3.7--3.7) & & 4.2 (3.6--4.4) \\
\end{tabular}
\caption{Numerical results of the star-by-star analysis of the cluster. The
parenthesized age ranges indicate the 90\% confidence limits.}
\label{tab_res_num}
\end{table}
\begin{table}
\begin{tabular}{ccc}
Quantity & $L$(He{\sc II} $\lambda$4686) & N{\sc III} $\lambda$4640/He{\sc II}
$\lambda$4686 \\
 & (10$^{35}$ erg s$^{-1}$) & \\
\hline
Observed & 3.0$\pm$0.2 & 0.17$\pm$0.01 \\
Model & 13$\pm$9 & 0.2$\pm$0.1 \\
\end{tabular}
\caption{WR line strengths, observed/predicted in the dereddened CAHA spectrum.}
\label{tab_wr}
\end{table}

\subsection{Mean IMF}
\label{meanimf}
\begin{figure}
\resizebox{\hsize}{!}{\includegraphics[angle=90]{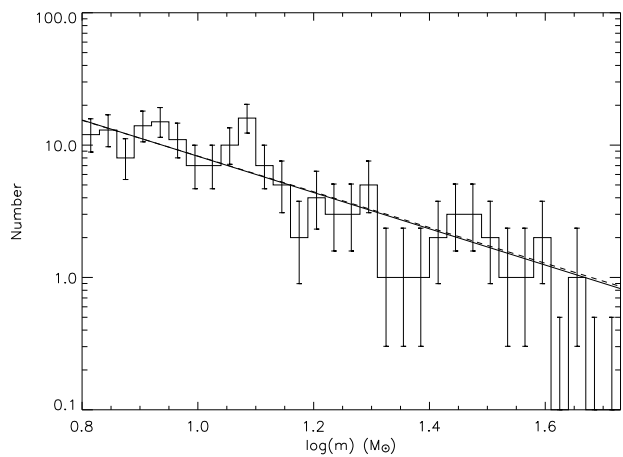}}
\caption{IMF histogram and fit. Full line: best fit. Dashed line: fit for the
Salpeter slope. The error bars are equivalent to 1$\sigma$ Gaussian ones, in
terms of likelihood, and must be interpreted this way: for a model that goes
through the extremity of an error bar, the observed value whose error bar is
being considered has a likelihood to occur of $\exp(-1/2)$ times the maximum
likelihood of the model at this abscissa.}
\label{fig_hist_imf}
\end{figure}

From the initial mass associated with each star, we computed the mean power-law
IMF of the cluster. More specifically, we constructed a histogram of the
logarithm of the initial masses, from $\log(m)=0.8$ to $\log(m)=1.76$ ($m=6.3$
to $m=58$ M$_{\sun}$), with a bin of 0.03 (equivalent to a factor of 1.07
between two successive bins), and fit it with an exponential law, assuming
Poissonian noise in each bin, knowing that a power-law IMF $dN/dm\propto
m^{-\alpha}$ can be translated into the law $dN/d\log(m)\propto
10^{(1-\alpha)\log(m)}$. The lower limit of the initial mass range was chosen
in order to ensure that the stars of the cluster belonging to a given bin were
all detected, and the upper limit corresponds to the highest possible initial
mass at the age and metallicity of the cluster. The result is shown in
Fig.~\ref{fig_hist_imf}. We found an IMF slope $\alpha=2.37\pm 0.16$, and
retained the compatible Salpeter slope, resulting in an inferred IMF
$dN/dm=(2680\pm 210)\,m^{-2.35}$. The corresponding total initial mass
integrated in the range $1\leq m\leq 58$ M$_{\sun}$ is $M_{\rm tot}=5800\pm
500$ M$_{\sun}$.

\section{Discussion}
\label{discuss}
The most massive stars of an OB association are also the most influential ones,
though the least numerous, on the spectrum of this cluster. In moderately
massive clusters, their small number is subject to significant fluctuations
around the mean IMF, causing large variations in the observable (near UV and
optical) and Lyman continuum spectral ranges. For instance, in Section
\ref{res_class}, we saw the significance of the presence of only 0.8 BSG in the
analytical model of \object{NGC 588} cluster upon the overall properties
derived from this model. We now discuss the effects of the real sampling of the
IMF in various mass ranges associated to different kinds of massive stars, and
more generally in the mass range covered by the most luminous stars. Then, we
present the results of a simulation of the effect of discrete random IMF
sampling, in the whole initial mass range, over $W({\rm H}\beta)$.

\subsection{Mean and observed numbers of different kinds of massive stars}
\label{massivekinds}
From the mean IMF computed in Section~\ref{meanimf}, we derived the mean
expected numbers of several kinds of massive stars susceptible to change
significantly the spectral properties of the cluster. Here we present these
results and discuss them.

\subsubsection{Blue supergiants}
\label{bsgeff}
According to our stellar models, at 4.2 Myr and for Z=0.008, BSGs (as defined
in \S~\ref{res_class}) are stars with initial masses ranging from 44 to 55
M$_{\sun}$. The number expectancy, derived from the integration, between these
two limits, of the mean power-law IMF computed in \S~\ref{meanimf}, is
${\mathcal N}({\rm BSG})=3.0$. The associated Poisson probability to observe no
BSG, as is our case, is $\exp({\mathcal N}({\rm BSG}))=5$\%. We added an
artificial continuous population composed by this kind of stars, weighted by
the Salpeter IMF slope, to the actual stellar content of the cluster.
Independently of the optical absolute luminosity -- which would be
automatically fit in a classical modeling of the cluster -- the most striking
change in the computed spectral properties is, as expected, the large drop of
$W({\rm H}\beta)$: in presence of the BSG population, the latter quantity would
be only 133 {\AA}, instead of 342 {\AA}.

\subsubsection{Wolf-Rayet stars}
\label{wreff}
We computed the expected number of the different WR stars (WNL, WNE, WCL, WCE,
WO), derived from the mean IMF of \S~\ref{meanimf}; we found ${\mathcal N}({\rm
WNL})=0.15$, ${\mathcal N}({\rm WNE})=0.02$, ${\mathcal N}({\rm WCL})=0.02$, 
${\mathcal N}({\rm WCE})=0.01$ and $ {\mathcal N}({\rm WO})=0.29$, according to
the star classification of Starburst99. The resulting Poisson likelihood of the
observed WR content of the cluster is 0.7\%, a small value that is however much
more satisfactory than the 0.04\% likelihood derived from the analytical model.
Furthermore, we simulated a spectrum obtained by replacing the two WR stars of
the cluster with an analytical, complete population of 0.51 WN star following
the Salpeter IMF. This population was expected to produce less luminous but
possibly harder or softer radiation than the two actually detected WR, as the
WR branch of the HR diagram (Fig.~\ref{hrdiag}) spans a significant range of
effective temperatures. We found that whereas $W({\rm H}\beta)$ is smaller with
the modified stellar population, the latter produces harder radiation than the
``true'' one in the Lyman continuum range, especially below the threshold of
ionization of He$^+$, as summarized in Table~\ref{tab_wrmodif}.
\begin{table*}
\begin{tabular}{lccccc}
BSG content & real & IMF & real & IMF$^a$ & 2.8 Myr$^{a,b}$ \\
WR content & real & real & IMF & IMF$^a$ & 2.8 Myr$^{a,b}$ \\
MS content & real & real & real & IMF$^a$ & 2.8 Myr$^{a,b}$ \\
\hline
$W({\rm H}\beta)$ ({\AA})& 342 & 133 & 284 & 80 & 247 \\
$Q({\rm H}^0)$ ($10^{50}$ s$^{-1}$)& 2.14 & 2.16 & 1.39 & 0.54 & 1.60 \\
$Q({\rm He}^0)/Q({\rm H}^0)$& 0.13 & 0.12 & 0.14 & 0.11 & 0.14 \\
$10^8$ $Q({\rm He}^+)/Q({\rm H}^0)$& 9.3 & 8.0 & 2.8$\times$10$^4$ & 8.8 & 130\\
$Q({\rm O}^+)/Q({\rm H}^0)$& 0.020 & 0.020 & 0.024 & 0.011 & 0.017 \\
\end{tabular}
\caption{Some spectral properties derived from the observed stellar content
(2nd column) and from the one with BSG, WR and whole population replaced with
the one predicted by the fitted power-law IMF (3rd, 4th and 5th columns,
respectively). $Q(X)$ designates the photon rate able to ionize the ion ``X''
population of a nebula. $^a$The 5th and 6th columns correspond to analytical
models, whose total mass here reproduce the observed F439W luminosity.
$^b$Best-fit model of the canonical analysis.}
\label{tab_wrmodif}
\end{table*}

\subsection{On the importance of the few most massive stars}
\label{fewmassive}
In \S~\ref{bsgeff} and \ref{wreff}, we saw the significance of fluctuations in
the population of some characteristic kinds of massive stars upon the spectral
properties of a cluster as moderately massive as the one of \object{NGC 588}.
This issue can be generalized to all sorts of very massive stars, that can
radiate intensely in the optical and/or ionizing spectral range and whose
numbers are subject to the largest relative statistical fluctuations.

\begin{figure}
\resizebox{\hsize}{!}{\includegraphics[angle=90]{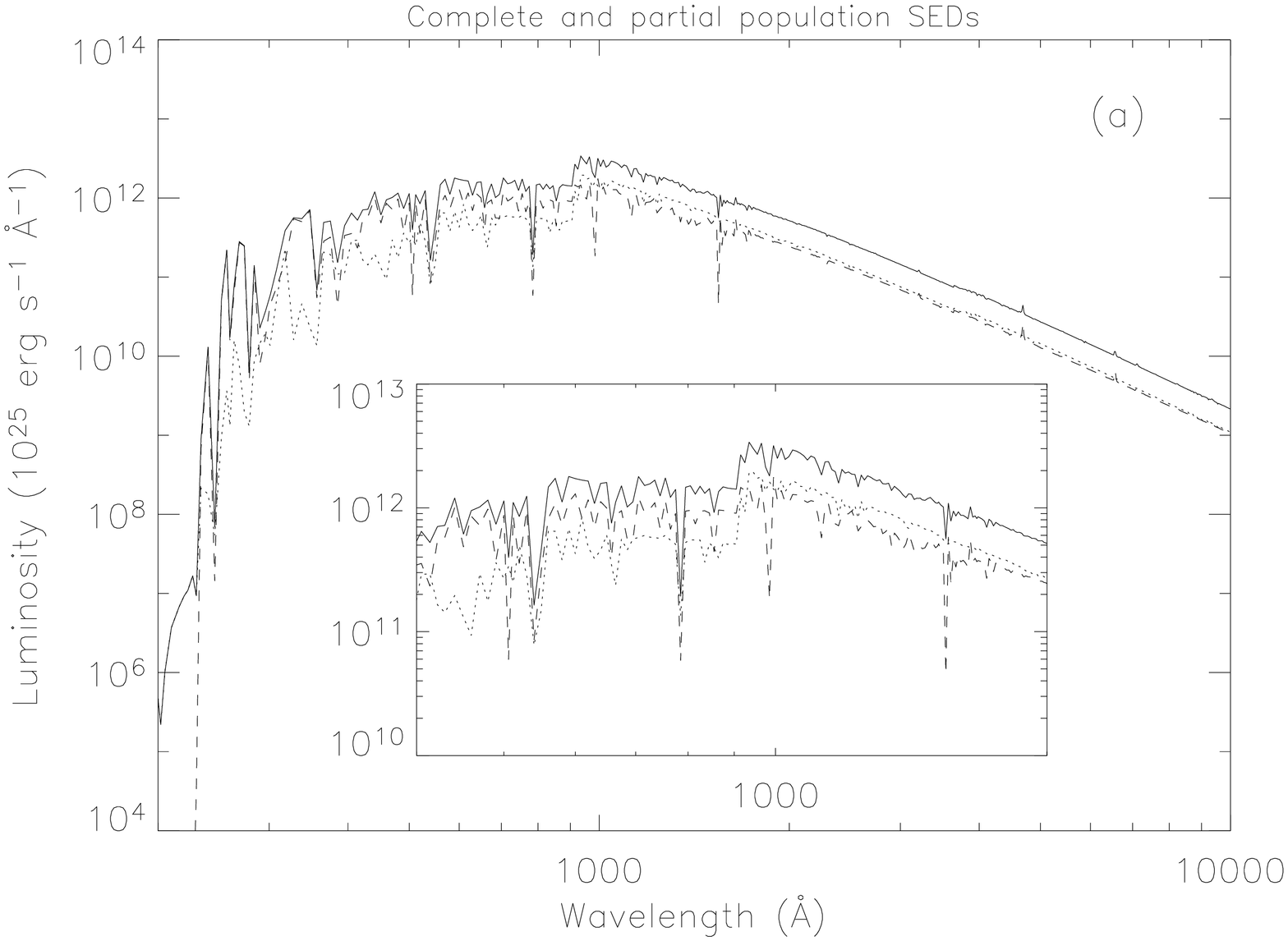}}
\resizebox{\hsize}{!}{\includegraphics[angle=90]{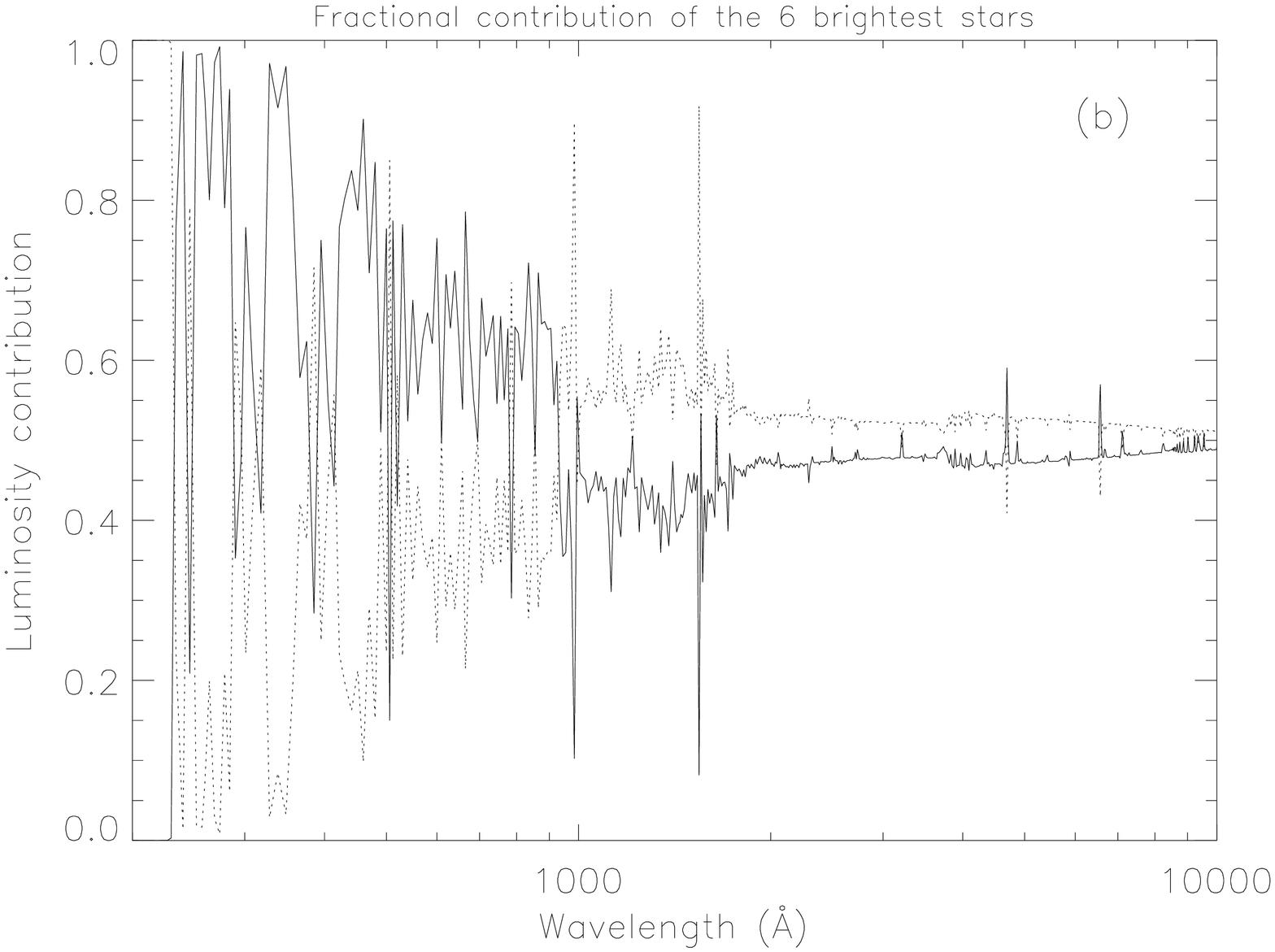}}
\caption{Top: SEDs of the whole cluster (full line), of the 6 brightest stars
(dashed line) and of the remaining 167 stars (dotted line). Bottom: ratio of
the brightest 6 stars SED (full line) and of the remaining 167 star SED (dotted
line) to the total SED.}
\label{sed6stars}
\end{figure}
\begin{figure}
\resizebox{\hsize}{!}{\includegraphics{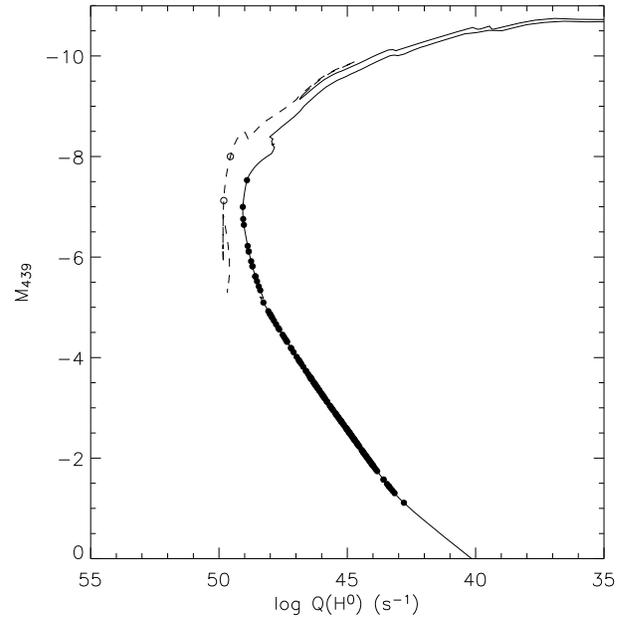}}
\caption{$(\log Q({\rm H}^0),M_{439})$ diagram of the detected stars along the
the theoretical isochrone. The WR branch is the dashed part of the curve.}
\label{magbqhdiag}
\end{figure}
\begin{figure}
\resizebox{\hsize}{!}{\includegraphics[angle=90]{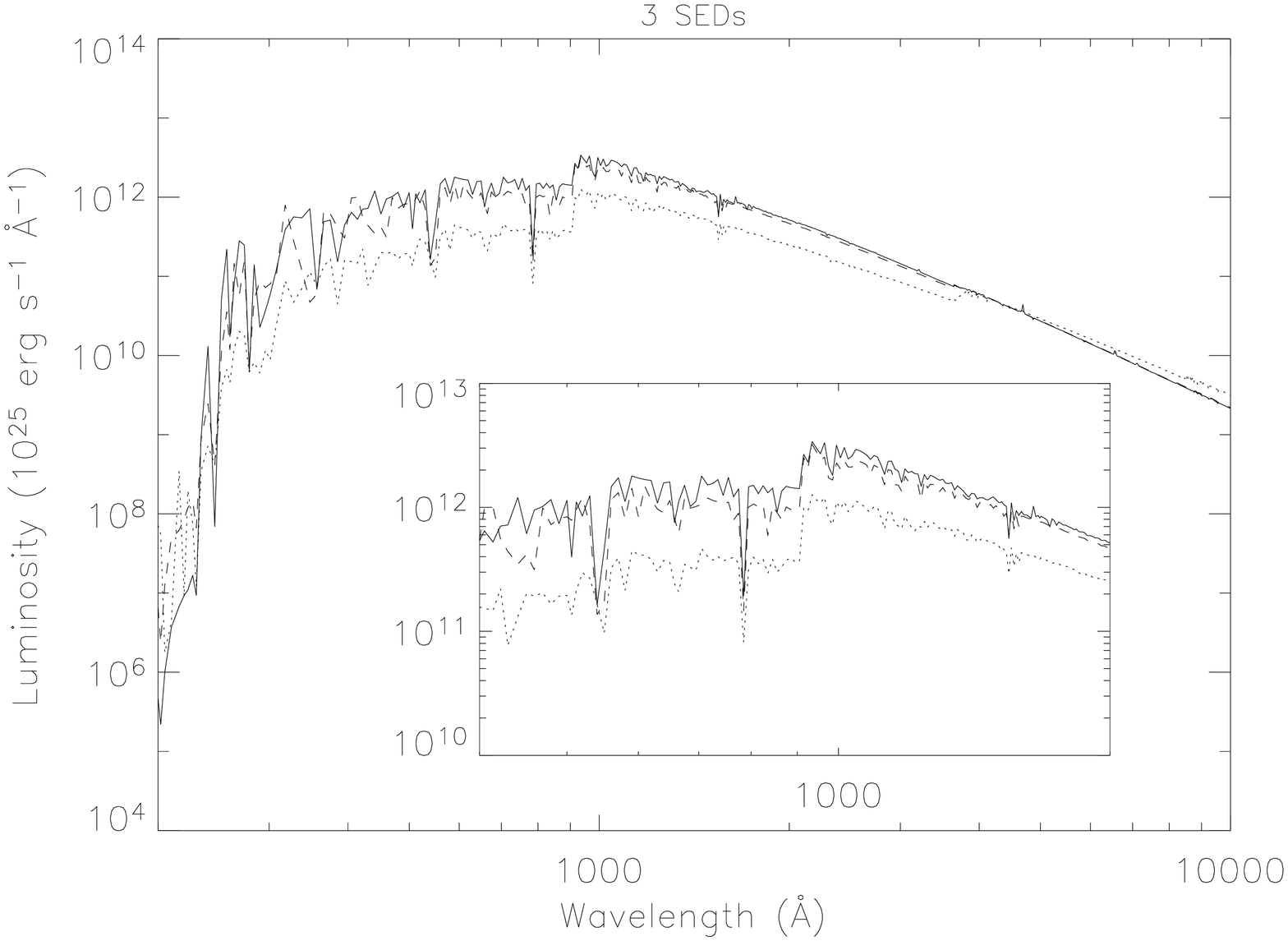}}
\caption{Star-by-star constructed SED (full lines), and SEDs of the analytical
populations, for Z=0.008 and $m_{\rm up}=120$ M$_{\sun}$, at 2.8 Myr (dashed
line) and 4.2 Myr (dotted line).}
\label{fig_3seds}
\end{figure}

In Fig.~\ref{sed6stars} are plotted the SEDs of the whole cluster, of the 6
brightest stars (in filter F439W) and of the remaining 167 stars, as well as
the fractional contribution of the 6 brightest stars to the total SED. It is
evident that these 6 stars are very influential on the total spectrum, whether
at observable wavelengths (where they are responsible for about half the total
flux) or in the high-energy range, where they generate approximately two thirds
of the hydrogen-ionizing photons, even though they constitute only about 1/30
of the whole set of detected stars. These 6 stars, which also are the most
massive ones, are specifically the stars situated in the temperature-luminosity
zone that varies significantly within the first few millions of years of the
cluster, as one can see in Fig.~\ref{hrdiag}. Fig.~\ref{magbqhdiag} shows the
$(M_{439},\log Q({\rm H}^0))$ diagram of the detected stars. The most striking
feature is the quasi-vacuousness of the curve in the region of the 6 dominating
stars, in particular the BSG branch that drops towards negligible Lyman
continuum at high band F439W luminosities. If the IMF of the cluster were
complete, as assumed in a classical model of stellar population, then the main
spectral diagnostic, $W({\rm H}\beta)$, would be significantly lower than what
we observe at the age and metallicity we derived from the star-by-star
approach. This is illustrated in Fig.~\ref{fig_3seds}, where the SED found with
the classical approach is also shown. In this figure, we can appraise the
significance of the drop of the Lyman-to-optical continuum ratio, as well as
the increase of the magnitude of the Balmer jump and the decrease of the global
slope of the SED at observable wavelengths, when passing from the star-by-star
SED to the one of the analytical model.

\subsection{Approximate expected uncertainties and Monte-Carlo simulation}
In Section~\ref{massivekinds} and \ref{fewmassive}, we saw the significant
consequences of the deviation of the actual (discrete and statistically
fluctuating) IMF of \object{NGC 588} from the mean and continuously sampled
analytical one. In this cluster of $\approx$6000 M$_{\sun}$ (integrated in the
interval of individual initial stellar masses 1--58 M$_{\sun}$), the stars
massive enough to rule its spectral properties, whether observable ($W({\rm
H}\beta)$, Balmer jump, etc.) or not (e.g., hardness of the Lyman continuum),
are only 6, which can easily explain the encountered issues.

\begin{figure}
\resizebox{\hsize}{!}{\includegraphics[angle=90]{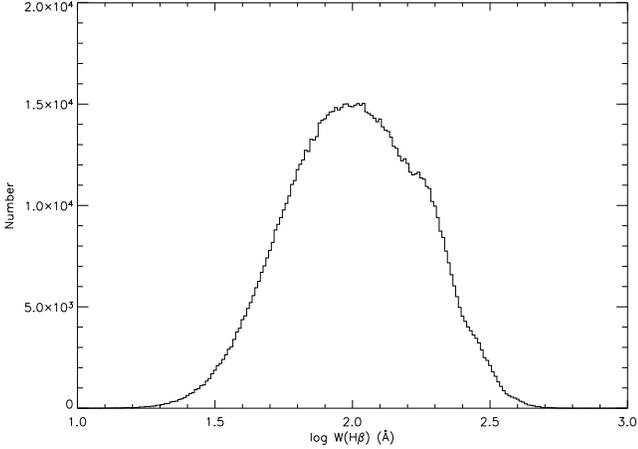}}
\caption{Histogram of $\log W({\rm H}\beta)$ resulting from the 10$^6$
Monte-Carlo discrete IMF realizations.}
\label{hist_logewhb}
\end{figure}

In order to estimate quantitatively the uncertainties related to the
fluctuations of massive stellar populations of same age, metallicity and IMF as
the ones of \object{NGC 588}, we performed a simple Monte-Carlo simulation
where the mean IMF of \S~\ref{meanimf} was divided into 0.01 M$_{\sun}$ wide
mass bins, in which 1977 stars (1977 being the number of stars integrated in
the mean IMF) were randomly distributed, according to the probability of each
bin to be filled with a given star, this probability being given by the mean
analytical IMF. We performed 10$^6$ such Monte-Carlo realizations. This model
assumes the stars of the cluster to have random masses, uncorrelated to each
other (a simplification that may be corrected when the actual properties of
star formation in clusters are better known), and following a likelihood
function given by the mean IMF. For each random distribution, we calculated the
resulting $W({\rm H}\beta)$; we found $\log(W({\rm H}\beta))$ to follow a
roughly Gaussian distribution with standard deviation 0.24 dex \citep[as also
obtained analytically by][]{cvl02}. The histogram of $\log(W({\rm H}\beta))$ is
shown in Fig. \ref{hist_logewhb}. According to the simulation, the
maximum-likelihood value of $W({\rm H}\beta)$ is 68 {\AA}, and the 90\%
likelihood interval for this quantity is as wide as 40--244 {\AA}. The observed
value $W({\rm H}\beta)=330\pm30$ {\AA} is marginal, the likelihood of realizing
a value greater than or equal to this one being $\approx$1\%. This likelihood
is of same order of magnitude as the 0.4\% one to observe, as we do, no BSG and
2 WNL stars, given the mean IMF of the cluster.

Using the same kind of simulations, we established that, for the same IMF slope
and cutoffs, the fluctuation of $W({\rm H}\beta)$ would still be $\approx$20\%
for a cluster as massive as $50\times10^3$ M$_{\sun}$.

\section{Conclusions}

In this work we have gathered the widest range of data available for the
stellar cluster ionizing the giant H{\sc II} region \object{NGC 588}, both
imaging and spectroscopy, covering a wide wavelength range from the ultraviolet
to the far red.

In the first part (Section~\ref{obsdat} and \ref{classi}), we showed the
importance of endeavoring to obtain and process as numerous and accurate
observable data as possible, in order to constrain models of the most common
form of OB associations within the limits of the relevance of these models,
rather than within the measurement errors and uncertainties. We discussed the
problems encountered when fitting overall integrated nebular and stellar
parameters, by means of a standard evolutionary population synthesis approach.
In particular, by assuming an analytical IMF, the best fit model predicts the
presence of BSGs simultaneous to the required (since observed) existence of
WNs, and consequently predicts too a low value of $W({\rm H}\beta)$. This
failure can be imputed mainly to two physical issues: the still important
uncertainties in our knowledge of the evolution of massive stars \citep[as
reviewed by][]{m03}, and the effects associated with IMF sampling \citep[as
studied by][]{cvl02}. We explored the latter in Section~\ref{statmod} through
\ref{discuss}. In Section~\ref{photappr}, we achieved quite a robust model of
the cluster, and obtained good results. However, in this approach, we inferred
a cluster age 50\% higher than with the analytical approach, and a mass twice
as large. In Section~\ref{discuss}, we estimated the effects of departure of
the observed set of stars from the one derived from the full sampling of the
mean cluster IMF. In particular, we established that a diagnostic such as
$W({\rm H}\beta)$ is very sensitive to the BSG content of the cluster, and that
the hardness of the ionizing radiation, an unobservable spectral property that
plays an important role in photoionization models of nebulae, depends
significantly on the WR content. More generally, we assessed that the radiation
field of \object{NGC 588} is one among a large variety of possible SEDs for
such a cluster, and that this is easily explained by the following statements:
{\em the most massive stars dominate the flux of the cluster both in the
observable and Lyman-continuum wavelength ranges, show a wide variety of
spectral properties, and are subject to strong population fluctuations}.

We suggest three practical considerations that will help to better understand
and characterize the evolutionary stage of massive ionizing star clusters: (i)
A ``best fit model'' statement may not be very meaningful per se, and should
not make us complacent that it implies the best solution. (ii) When possible,
use color-magnitude diagrams to obtain the integrated properties of the
ionizing cluster, as opposed to assuming a perfectly sampled mass function.
(iii) Assess the importance of the few most massive stars on the overall SED of
the cluster, both at ionizing and at non-ionizing wavelengths; this can be
easily done by applying the lowest luminosity limit criteria established by
\cite{clp03} and \cite{cl04}. If possible, use Monte-Carlo population models,
provided they are relevant, and pending a more mature development of this
approach to population synthesis.

We also established that the initial mass distribution of the stars detected in
\object{NGC 588} is compatible with a Salpeter IMF, if it is treated as a
stochastic process.

Finally, in this work, we used the very common hypothesis of instantaneous
starburst. However, recent works, like the one of \cite{tps03}, show that star
formation in massive clusters spans time ranges such that the instantaneous
formation hypothesis may not be valid for them. This statement does not call
into question the significance of the disturbances caused by fluctuations in
the high mass end of the IMF, but it is an additional uncertainty that is worth
accounting for in further models of massive star clusters.

\begin{acknowledgements}

This work could not have been done without the contemporary direct access to a
variety of publicly available astronomical archives. It is based on
observations taken with the 3.5m telescope at the Centro Astron\'omico
Hispano-Alem\'an (CAHA, Calar Alto, Almer\'\i a), on observations taken with
the NASA/ESA Hubble Space Telescope, obtained at the Space Telescope Science
Institute, which is operated by the Association of Universities for Research in
Astronomy, Inc., under NASA contract NAS5-26555, on INES data from the IUE
satellite, and on data from the ING Archive taken with the JKT operated on the
island of La Palma by Isaac Newton Group of Telescopes in the Spanish
Observatorio del Roque de Los Muchachos of the Instituto de Astrof{\'\i}sica de
Canarias.

Funding was provided by French CNRS Programme National GALAXIES, by Spanish
grants AYA-2001-3939-C03-01, AYA-2001-2089, and AYA2001-2147-C02-01, by Mexican
grant CONACYT 36132-E, and by French-Spanish bi-lateral program
PICASSO/Acci\'on Integrada HF2000-0143.

We also thank Daniel Schaerer for his help on the WR bumps. We are also
grateful to Miguel Mas-Hesse and the referee of the present article for careful
reading and very useful comments and suggestions.

\end{acknowledgements}

\appendix{}
\section{Magnitude interpolation}
At given age and metallicity, the isochrone data available from our programs
consist in an array of physical parameters of model stars, each line containing
the following parameters for the current star: the initial mass $M$, the
effective temperature $T_{\rm eff}$, the bolometric luminosity $L_{\rm bol}$,
and the surface abundances, in particular the hydrogen one $X$. From these
parameters, we wanted to derive absolute magnitudes, to be compared to the
observations, from atmosphere models. In our case, four kinds of models were
available: \cite{lbc97}, \cite{phl01}, \cite{hm98} and black body. For each of
the first three kinds of models, whose selection for given physical parameters
was operated the same way as Starburst99 does, stellar fluxes are tabulated for
a fix array of wavelengths and for various effective temperatures; in the case
of Lejeune and Pauldrach models, the surface gravity $g$ is also present. When
one of these grids is selected, Starburst99 performs nearest research
interpolation along the $\log T_{\rm eff}$ axis (for Hillier grids) or in the
$(\log T_{\rm eff},\log g)$ plane (for Lejeune and Pauldrach grids) to select a
flux array, and normalizes the latter for its total to be equal to $L_{\rm
bol}$. This non-continuous interpolation method can be exploited to compute the
total spectrum of a model cluster, as the interpolation errors will tend to
cancel each other. However, the resulting isochrone curve in a
magnitude-magnitude of color-magnitude diagram will be irregular, with jumps
resulting from the sudden transition from an array of the exploited grid to
another. Here we suggest another form of interpolation to obtain more
accurately determined model magnitudes.

Let us first consider the case of the black-body emission law: the
wavelength-and-temperature-dependent flux is
\begin{equation}
{\mathcal B}_{\lambda}(\lambda,T)=\frac{2\pi h
c}{\lambda^5}\frac{1}{e^{hc/\lambda kT}-1}
\end{equation}
The change of variable $x=hc/\lambda k
T=1.44\times10^8/\lambda T$ then gives:
\begin{eqnarray}
\ln {\mathcal B}_{\lambda}(\lambda,T)&=&\ln\left(\frac{2\pi h
c}{\lambda^5}\right)-\ln(e^x-1) \\
&=&\ln\left(\frac{2\pi h
c}{\lambda^5}\right)-(x+\ln(1-e^{-x}))
\end{eqnarray}
This means that at a given wavelength $\lambda$, $\ln {\mathcal
B}_{\lambda}(\lambda,T)$ is a linear function of the temperature-dependent
variable $y=\ln(e^x-1)=x+\ln(1-e^{-x})$. This is the main idea underlying the
interpolation of magnitudes from the grids of model stellar spectra. Indeed,
the wavelengths covered by the HST filters F547M, F439W, F336W and F170W are
line-free and keep quite far from the continuum jumps (in particular, the
Balmer jump); hence at these wavelengths, we expect the stellar fluxes to
behave nearly as in the black-body case. Consequently, we may calculate the
flux predicted at wavelength $\lambda$, for the effective temperature $T_{\rm
eff}$ and the gravity $g$, with the fluxes available in the grid in use at the
two temperatures $T_1$ and $T_2$ that ``surround'' $T_{\rm eff}$, and at the
same logarithmic surface gravity $L=\log g$, using the following linear
interpolation formula:
\begin{eqnarray}
\ln F_{\lambda}(\lambda;T_{\rm eff},L)~=\hspace{118pt}&&\nonumber \\
\frac{(y_2-y)\ln F_{\lambda}(\lambda;T_1,L)+(y-y_1)\ln
F_{\lambda}(\lambda;T_2,L)}{y_2-y_1}&&
\label{eq_interp}
\end{eqnarray}
This formula was straightly applied to the grids of Hillier model spectra,
ignoring the $L$ variable which is undefined for these grids. For Lejeune and
Pauldrach grids, we also needed to interpolate the fluxes as functions of $L$.
This was done the following way:
\begin{itemize}
\item{when, for the considered temperature, $T_1$ or $T_2$, at least two values
of the surface gravity were available, the two ones closest to $g$, namely
$g_1$ and $g_2$, were used to perform linear interpolation as a function of
$L$:}
\begin{eqnarray}
\ln F_{\lambda}(\lambda;T_i,L)~=\hspace{131pt}&& \\
\frac{(L_2-L)\ln F_{\lambda}(\lambda;T_i,L_1)+(L-L_1)\ln
F_{\lambda}(\lambda;T_i,L_1)}{L_2-L_1}\nonumber
\end{eqnarray}
\item{when only one value surface gravity was available in the grid (a marginal
case that could occur with the Lejeune grids for the highest temperatures), its
corresponding array of fluxes was adopted ``as is'', i.e. neglecting the
dependence of the stellar flux on gravity.}
\end{itemize}
When using Lejeune or Pauldrach grids, the interpolation algorithm was finally
the following:
\begin{enumerate}
\item{in the grid, search for the two effective temperatures $T_1$ and $T_2$
nearest the temperature of interest $T_{\rm eff}$;}
\item{for each of both $T_i$, determine the two values $g_{i1}$ and $g_{i2}$ of
the surface gravity closest to $g$, and calculate the array of fluxes
$F_{\lambda}(\lambda;T_i,\log g)$ by linear interpolation in the ``segment''
$[\log g_{i1},\log g_{i2}]$. If actually only one value of the gravity $g_i$ is
available, then set $F_{\lambda}(\lambda;T_i,\log
g)=F_{\lambda}(\lambda;T_i,\log g_i)$;}
\item{at each wavelength, calculate $y$, $y_1$ and $y_2$, and apply formula
(\ref{eq_interp}) to determine the interpolated flux. Note that due to the fact
that the variable $x=1.44\times10^8/\lambda T$ could have values up to
$\approx800$, we used the formula $y=x+\ln(1-e^{-x})$ and not $y=\ln(e^x-1)$,
as the latter would have caused floating overflows when running our codes
($e^{800}\sim10^{347}$).}
\end{enumerate}
As we saw above, our interpolation method is meaningful for wavelength ranges
where the stellar fluxes behave nearly as black-body ones. However, it does not
apply at wavelengths of lines or near discontinuities (e.g., Lyman and Balmer
jumps). The interpolation proposed here is preferred to calculate smoother
running isochrone magnitudes, but does not involve a clear advantage to compute
model spectra of clusters.

\end{document}